\documentclass[man, natbib, apacite, noextraspace, 12pt, floatsintext]{apa6}

\usepackage{comment}
\usepackage{amsmath,amssymb}
\usepackage{bm}
\usepackage{graphicx}
\usepackage{mathtools}
\usepackage{makeidx}  
\usepackage{makecell}
\usepackage{url}
\usepackage{tikz}

\usepackage{array}
\usepackage{float}
\usepackage{booktabs, caption}
\usepackage{threeparttable}
\usepackage{pdflscape}
\usepackage{setspace}
\usepackage{longtable}

\newcommand{\rev}[1]{#1}

\newcommand{\etat}{\eta^T}
\newcommand{\etati}{\eta^{T}_i}
\newcommand{\bM}{\bm{M}}
\newcommand{\bx}{\bm{x}}
\newcommand{\yt}{Y^T}
\newcommand{\yti}{Y^T_i}
\newcommand{\yc}{Y^C}
\newcommand{\yci}{Y^C_i}
\newcommand{\Mt}{M^T}
\newcommand{\bMt}{\bm{M}^T}

\newcommand{\btheta}{\bm{\theta}}

\newcommand{\NA}{\mbox{\textsc{na}}}

\newcommand\independent{\protect\mathpalette{\protect\independenT}{\perp}}
\def\independenT#1#2{\mathrel{\rlap{$#1#2$}\mkern2mu{#1#2}}}

\title{Fully Latent Principal Stratification With Measurement Models}
\date{}

\shorttitle{Fully Latent Principal Stratification}

\author{Sooyong Lee$^{1}$ \\
Adam Sales$^{2}$ \\
Hyeon-Ah Kang$^{1}$ \\
Tiffany A. Whittaker$^{1}$}

\affiliation{$^{1}$The University of Texas at Austin \\
$^{2}$Worcester Polytechnic Institute}

\abstract{There is wide agreement on the importance of implementation data from randomized effectiveness studies in behavioral science; however, there are few methods available to incorporate these data into causal models, especially when they are multivariate or longitudinal, and interest is in low-dimensional summaries.  We introduce a framework for studying how treatment effects vary between subjects who implement an intervention differently, combining principal stratification with latent variable measurement models; since principal strata are latent in both treatment arms, we call it “fullylatent principal stratification” or FLPS. We describe FLPS models including item-response-theory measurement, show that they are feasible in a simulation study, and illustrate them in an analysis of hint usage from a randomized study of computerized mathematics tutors.}

\begin{document}

\maketitle



\setcounter{secnumdepth}{2}

\section{Introduction}

It is a truth universally acknowledged (and stipulated in funding agreements) that researchers conducting randomized controlled trials (RCTs) should gather data not just on subjects' baseline characteristics, treatment assignments, and outcomes, but also on the extent to which they actually implemented the intervention, or ``implementation fidelity.'' 
As \citeA[p.2]{carroll2007conceptual} argue, ``evaluation of implementation fidelity is important because this variable may not only moderate the relationship between an intervention and its outcomes, but its assessment may also prevent potentially
false conclusions from being drawn about an intervention's effectiveness.'' For instance, a null result may be due to an intervention's ineffectiveness, or to poor implementation.

Of course, implementation can vary from subject to subject in an experiment, and, in principle, researchers can exploit this variation to estimate the extent to which implementation fidelity, as described by \citeA[p.2]{carroll2007conceptual}, ``moderate[s] the relationship between an intervention and its outcomes,'' or to test the hypothesis that a null result is due to poor implementation (i.e., by showing that an intervention is effective for subjects who implement it properly). 
However, there are few methods available---and even fewer widely known---to actually accomplish these tasks.
While everyone agrees that measuring implementation fidelity is important, few researchers know how to use those measurements (or, at least, how to incorporate them into formal causal models). 

The recent profusion of educational technology (EdTech) products---and RCTs evaluating them---has made the situation even more challenging, but also more promising. 
EdTech programs automatically gather log data on every action that each user takes, yielding huge, rich datasets describing implementation in fine detail.
However, log data are messy, complex, and impossible to make sense of without a suitable model (which itself must be sufficiently complex).
Surely, this resource could be harnessed to better understand the varying effectiveness of EdTech products, but as yet it is unclear how (see \citeA{sales2021student} for some suggestions). 

Causal modeling of implementation fidelity involves other serious methodological challenges, chief among them is the fact that implementation is endogenous---that is, subjects who implement an intervention are likely to be different at baseline from those who do not. 
In the framework of principal stratification \cite<PS;>{frangakis} 
(as applied to implementation fidelity), subjects in an experiment are divided into ``principal strata'' based on how they would potentially implement an intervention if assigned to the treatment condition. 
For instance, given a two-component intervention, there could be four principal strata, composed of subjects who would, if assigned to treatment, implement each component by itself, both components, or neither. \label{oneTwoWayNoncomp}
\rev{(If subjects assigned to control might have access to the treatment, then subjects' potential implementation if assigned to control must also be included as principal strata.)}
Then, researchers may estimate ``principal effects'' or average effects for subjects in each principal stratum.  
Principal stratum membership is only observed for subjects assigned to treatment, but is latent for subjects assigned to control---their potential implementation must be modeled and/or imputed using covariates measured at baseline. (Actually, PS applies to a much broader set of problems than described here; see \citeA{page2015principal,mealli2012refreshing} for reviews.)

PS, then, is an appropriate framework for assessing the role of implementation fidelity in an intervention's effect, as long as implementation is observed, or measured without error, in the treatment group. However, it is increasingly common that implementation must be modeled, or is measured with error. For instance, \citeA{vanacore_ottmar_liu_sales_2023} uses a normal mixture model to classify classrooms as high or low implementation settings in an RCT studying educational technologies.
\citeA{aoas} showed that when implementation is measured with error, classic PS models ignoring measurement error can fit poorly and yield misleading conclusions, while including an explicit measurement model into PS can lead to better estimates.

This paper will generalize and build off of those results, describing an alternative framework, called ``fully-latent principal stratification'' (FLPS), that generalizes PS to scenarios in which implementation in the treatment group is measured as a latent construct. That is, in FLPS, implementation fidelity is not directly observed in either the treatment or the control group---hence, ``fully latent''---though relevant indicators of implementation are observed for subjects assigned to treatment. 
FLPS models seek to estimate the relationship between implementation fidelity (or, really, any aspect of program implementation) and program effects in two steps: first, use a latent-variable model to measure implementation in the treatment group of an RCT, then use classical PS methods (suitably adopted to accommodate measurement error) to estimate principal effects.

After reviewing classical PS (Section \ref{sec:classicPS}), this paper will introduce the FLPS framework, define its estimands, describe identification conditions and a general estimation strategy, and give some parametric examples (Section \ref{sec:flps}).  Section \ref{sec:simulationDesign} presents \rev{a proof-of-concept\label{proofOfConceptIntro} simulation study of FLPS models based on four models from item response theory (IRT), showing that these FLPS models are feasible.} Section \ref{sec:realData} illustrates FLPS using data from an RCT comparing the efficacy of two EdTech programs and Section \ref{sec:discussion} concludes.

\section{Classical Principal Stratification}\label{sec:classicPS}

Suppose an RCT is designed to measure the treatment effects in an outcome $Y$ (e.g., math achievement) between treatment assignments (e.g., assignment to use an EdTech product versus business as usual).
Let $Z_i  \in \{0,1 \}$ indicate subject $i$’s treatment assignment, for $i=1,...,N$, and let $Y_i$ be the outcome of interest. Then, define two “potential outcomes” \cite{neyman,rubin} for subject $i$: $\yci$ is the value of $Y_i$ that would be measured were subject $i$ to be assigned to control, $Z_i = 0$, and $\yti$ is the value of $Y_i$ that would be measured were subject $i$ to be assigned to treatment. Although only one of subject $i$’s two potential outcomes is ever observed—$\yti$ if $Z_i = 1$ or $\yci$ if $Z_i = 0$—they are both defined for every subject $i$.

\sloppy
Define subject $i$’s individual treatment effect as $\tau_i \equiv \yti - \yci$. Typically, the goal of an RCT is to estimate the average treatment effect (ATE): $\mathbb{E}[\tau] = \mathbb{E}[\yt] - \mathbb{E}[\yc]$, where the expectation is taken over the sample of subjects in the experiment or a suitable superpopulation. Researchers may also estimate $\mathbb{E}[\tau|X=x] = \allowbreak \mathbb{E}[\yt|X=x] - \allowbreak \mathbb{E}[\yc|X=x]$, the ATE for the subset of subjects for whom a pre-treatment covariate $X$ takes the particular value $x$.

Let $M$ be a variable defined after treatment assignment, which may itself be affected by the treatment. In an RCT evaluating an EdTech product, $M$ could be derived from the software’s log data---say, the number of problems a student worked. Unlike subgroup effects for a pre-treatment variable $X$, estimands of  $\mathbb{E}[\tau|M=m]$ are not causally meaningful.

Principal stratification uses the structure of potential outcomes to define an alternative estimand. Like outcomes $Y$, $M$ has potential values $M^C$ and $\Mt$, corresponding to the values of $M$ that would be observed in the control and treatment conditions, respectively. When $M$ itself is a feature of the intervention, $M$ is undefined in the control group, so only $\Mt$ is defined. Like $\yt$, $\Mt$ is only observed for subjects in the treatment group, but is defined for every subject; it is a pre-treatment covariate that happens to be missing for subjects in the control group. Therefore, we can define ``principal effects'' for subgroups based on $\Mt$ (instead of $M$) as:
\begin{equation}\label{eq:prinEff}
\mathbb{E}[\tau|\Mt=m] = \mathbb{E}[\yt|\Mt=m] -\mathbb{E}[\yc|\Mt=m].
\end{equation}
This is the ATE for the subset of subjects for whom, were they assigned to the treatment condition, $\Mt$ would take the value $m$. In other words, the principal effect is a special type of subgroup or moderation effect—the effect of treatment assignment among subjects who would implement the intervention in a particular way.

For subjects in the treatment group (i.e. with $Z=1$), $\yt = Y$ and $\Mt = M$, so estimating the first term of \eqref{eq:prinEff} is straightforward.
Formally, we will focus on simple randomized experiments and assume
\begin{equation}\label{eq:ignorability}
\{\yt,\yc,\Mt,M^C,\bm{x}\}\independent Z.
\end{equation}
Under the assumption \eqref{eq:ignorability},
$
f(Y|Z=1,\Mt=m)
=f(\yt|\Mt=m)
$, 
and hence $\mathbb{E}[Y|Z=1,\Mt=m]=\mathbb{E}[\yt|\Mt=m]$.
Since whenever $Z=1$, $M=\Mt$, $\mathbb{E}[\yt | \Mt=m] = \mathbb{E}[Y | M=m,Z=1]$.
Similarly, $f(Y|Z=0,\Mt=m)=f(\yc|\Mt=m)$ and $\mathbb{E}[Y|Z=0,\Mt=m]=\mathbb{E}[\yc|\Mt=m]$.

The second term of \eqref{eq:prinEff} is trickier since $\Mt$ is missing whenever $\yc$ is observed (i.e., for members of the control group). The most common approach relies on parametric mixture modeling \cite<e.g.,>[] {gallop2009mediation,mattei2013exploiting,feller2016compared}. The distribution of $\yc$ conditional on a vector of covariates $\boldsymbol{x}$, can be decomposed as:
\begin{align}\label{eq:classicMixture}
f(\yc|\boldsymbol{x}) &=\sum f(\Mt|\boldsymbol{x})f(\yc|\Mt,\boldsymbol{x}), \text{ or} \nonumber \\ &= \int f(\yc | \Mt, \boldsymbol{x})f(\Mt|\boldsymbol{x}) d \Mt,
\end{align}
depending on whether $\Mt$ is discrete/categorical or continuous, with the sum or integral taken over the support of $\Mt$. Since both $\Mt$ and $\boldsymbol{x}$ are defined before treatment assignment, the relationship between them does not depend on the treatment assignment. Principal effects are calculated in terms of the parameters of the distributions of $\yc$, $\yt$, and $\Mt$, which themselves are typically estimated concurrently using the maximum likelihood or Bayesian Markov Chain Monte Carlo (MCMC) techniques.

The role that an intermediate variable $M$ plays in PS is fundamentally different from the role in traditional mediation analysis. In PS, $M$ is not a causal agent—the “effect of $M$” is never estimated. Instead, it is a realization of a subject’s baseline attribute, $\Mt$. The principal effect is a special type of subgroup or moderation effect—the effect of treatment assignment among subjects who would implement the intervention in a particular way.

\section{Fully Latent Principal Stratification}\label{sec:flps}

In the classical PS framework, $\Mt$---the intermediate variable that defines stratification---is a low-dimensional measurement without error.
Classical PS models can incorporate multiple measurements of $\Mt$ per subject by aggregating measurements, such as with the sample mean $\overline{\Mt}$, and using this aggregate as a unidimensional intermediate variable and stratifying on its potential values ($\overline{\Mt}_i$). This approach, however, ignores measurement error in the aggregate---a problem that is exacerbated when the number of measurements varies between individuals, leading to differential measurement error---in addition to other relevant aspects of the measurement structure.

For instance, we later introduce an RCT that evaluates computerized tutoring programs. In this example, $\Mt_{ij}=1$ if subject $i$ would receive feedback on problem $j$ and 0 otherwise. If $\overline{\Mt}_i$ is taken as an estimate of the probability of subject $i$ receiving feedback, then $Var(\overline{\Mt}_i)$ is inversely proportional to the number of problems $i$ has worked on, which could vary considerably between students. Moreover, students who worked on harder problems may have higher $\overline{\Mt}_i$ than their peers. Hence, $\overline{\Mt}_i$ incorporates information both on the number of problems worked and their average difficulty---aspects of students' implementation that might not be of direct interest in a particular analysis.

\subsection{FLPS}\label{sec:flpsIntro}

As an alternative, “Fully Latent PS” or “FLPS” extends the classical PS to model implementation data, including several measurements of an underlying construct of interest; denote these measurements as $\bMt_i$ for subjects $i$ assigned to the treatment condition.
FLPS incorporates the measurement process into PS, specifying a distribution for the measurements, $f(\bMt|\etat)$, where $\etat$ is a subject-level latent variable measured by the measurement outcomes. The $T$ superscript of $\etat$ is analogous to the superscripts of $\Mt$ and $\yt$; measurements $\boldsymbol{M}$ are only available for subjects in the treatment group, while $\etat$ is well-defined for all subjects in the experiment. It measures an aspect of subjects' potential implementation---how they would implement the intervention if assigned to the treatment condition.
\label{etaExamples}\rev{For instance, in \citeA{aoas}, $\etat$ measured a subject's propensity to master a set of attempted skills in a section of an Algebra I curriculum; in \citeA{gamers} it measured a subject's propensity to misuse an educational technology product; and in our analysis in Section \ref{sec:realData}, it measures a subject's propensity to receive help or feedback while working on practice problems. }
The causal estimand in FLPS is $\tau(\etat) \equiv \mathbb{E}[\tau|\etat]=\mathbb{E}[\yt|\etat]-\mathbb{E}[\yc|\etat]$, the ATE for subjects who would, if assigned to treatment, implement the interventions as $\etat$.

In FLPS, measurements $\bMt$ are considered indicators of latent construct $\etat$; $\bMt$ is never imputed for subjects with $Z=0$. Going forward, although $\bM_i$ is only measured when $Z_i=1$, we will drop the $T$ superscript and refer only to $\bM$.

In this paper, we will focus on the common scenario in education in which each measurement corresponds to a specific item $j=1,\dots,J$.
For instance, in \citeA{aoas}, the items were units from the curriculum of an educational computer application, so $\Mt_{ij}$ recorded whether student $i$ would master the material in unit $j$, and in the example in Section \ref{sec:realData}, the items are practice problems and $M_{ij}$ measures the amount of feedback student $i$ received while working on practice problem $j$.
In these examples, most students do not have measurements for every item---for instance, most students did not work on every available practice problem.
When no measurement is available for student $i$ on item $j$, we will write $\bM_{ij}=\NA$.
Let $\mathcal{J}_i=\{j: \bM_{ij}\ne\NA\}\subseteq [1,J]$, the set of items $j$ for which $i$ has a measurement.

\label{conditionalIndepence}
We will also assume that all information in $\boldsymbol{M}_i$ relevant to 
$Y$ 
is captured in $\etat$: 
\begin{equation}
        Y
        \independent \boldsymbol{M}|\etat,Z=1,\boldsymbol{x}\label{eq:independenceY}
\end{equation}
\rev{Assumption \eqref{eq:independenceY} requires some scrutiny. For instance, we will describe models in which $\etat$ is unidimensional; if, in reality, it is multidimensional, and potential outcomes depend on the multiple dimensions, then \eqref{eq:independenceY} will be violated. \citeA{aoas} demonstrates a test for multidimensionality in the FLPS context. }

Unlike the classical PS intermediate variable $\Mt$, latent variables, such as $\etat$, are not observed in either the treatment or control group of the study. Thus, distributions for both the treated and control potential outcomes will follow mixture distributions. Say $\etat$ is continuous; then, under assumption 
\eqref{eq:independenceY},
\begin{align}
   f(Y,\bm{M}|Z=1,\bm{x})&=
    \int f(Y,\bm{M},\etat|Z=1,\bm{x}) d\etat \nonumber\\
    &=\int f(Y|Z=1,\bm{x},\bm{M},\etat)f(\bm{M},\etat|Z=1,\bm{x})d\etat \nonumber\\
    &=\int f(Y|Z=1,\bm{x},\etat)f(\bM,\etat|\bm{x}) d\etat &\mbox{by \eqref{eq:ignorability}, \eqref{eq:independenceY}} \nonumber\\
    &=\int f(Y|Z=1,\etat,\bm{x})f(\bM|\etat,\bm{x})f(\etat|\bm{x})d\etat \label{eq:flpsMixtureT}
\end{align}
where $f(\etat|\boldsymbol{x})$ is a model for $\etat$ as a function of covariates, analogous to $f(\Mt|\boldsymbol{x})$ in \eqref{eq:classicMixture}.

The mixture distribution for control potential outcomes is
\begin{equation}\label{eq:flpsMixtureC}
    f(Y|Z=0,\bm{x})=\int f(Y |Z=0, \etat, \boldsymbol{x})f(\etat|\boldsymbol{x}) d \etat,
\end{equation}
that is, \eqref{eq:classicMixture} with $\Mt$ replaced by $\etat$. Although $\etat$ is unobserved in both the treatment and control groups, the data differ markedly between the two groups: the model for $\etat$ in the treatment group includes both measurements $\boldsymbol{M}$ and covariates, whereas, in the control group, only covariates are available.

\subsection{Bayesian Inference}\label{sec:estimation}

\rev{Let $\btheta$ be a generic vector of parameters for models \eqref{eq:flpsMixtureT} and \eqref{eq:flpsMixtureC}, with prior distribution $f(\btheta)$, and let $\mathcal{T}=\{i:Z_i=1\}$ and $\mathcal{C}=\{i:Z_i=0\}$ be the sets of indices for the treatment and control groups, respectively.}
The posterior probability density of $\btheta$ can be estimated via the following relation:
\begin{equation}\label{eq:posterior}
\begin{split}
f(\btheta|\bm{Y},\bm{Z},\boldsymbol{x}&,\boldsymbol{M}_\mathcal{T})   \\
\propto f(\btheta)&f(\bm{Y},\boldsymbol{M}_\mathcal{T}|\bm{Z},\bm{x},\btheta)\\
 = f(\btheta)&\prod_{i\in\mathcal{T}}\int f(Y_i|Z_i=1,\etati, \boldsymbol{x}_i, \btheta) f(\boldsymbol{M}_i | \etati, \bx_i,\btheta) f(\etati|\boldsymbol{x}_i, \btheta) d\etati    \\
\times&\prod_{i\in\mathcal{C}} \int f(Y_i | Z_i=0, \etati, \boldsymbol{x}_i, \btheta) f(\etati|\boldsymbol{x}_i, \btheta) d\etati,
\end{split}
\end{equation}
where the subscript $\mathcal{T}$ on $\bM$ 
denotes that these are only available for subjects assigned to the treatment condition. 


The integrals in \eqref{eq:posterior} are rarely tractable; instead, sampling can be used to evaluate the posterior distribution within MCMC. This study applies No-U-Turn sampling, which is the default sampler of Stan \cite{hoffman2014no}. The sampler is an efficient version of the Hamiltonian Monte Carlo sampler and has been shown to be computationally efficient for estimating the correlated parameters ~\cite{hoffman2014no}.

\subsection{\rev{Parametric FLPS Models}}\label{sec:models}

The FLPS framework introduced in the previous two sections, \ref{sec:flpsIntro} and \ref{sec:estimation}, encompass a broad range of models.
However, for the remainder of the paper, we will focus on a relatively narrow set of models, characterized by unidimensional measurement models from the item response theory and linear-normal models for $\etat$ and $Y$.
While the models we will discuss will include a variety of measurement submodels $f(\bm{M}|\etat,\btheta)$, they will all use identical specifications for submodels $f(\etat|\bm{x},\btheta)$ and $f(Y|Z,\bx,\etat,\btheta)$.

The measurement models we will discuss, drawn from the item response theory, are characterized by an item parameter $\bm{\zeta}_j$, which may be a vector, associated with measurements from each unique item students work on, $M_j$, and a scalar subject-level parameter $\etati$.
We assume local independence, that conditional on these parameters, all of subject $i$'s measurements are mutually independent, or, for $j\ne j'\in \mathcal{J}_i$,
$M_{ij}\independent M_{ij'}|\bm{\zeta},\etati$,
where $\bm{\zeta}$ (without the subscript) is a vector including item-parameters for all items.
Then the measurement model is completely specified by a model for each measurement occasion $M_{ij}$.
For binary measurements, $M_{ij}$, the simplest of those we will consider is the Rasch model \cite{rasch1960studies}
\begin{equation}\label{eq:rasch}
\phi_{ij}\equiv Pr(M_{ij}=1|\etati,d_j)=\mathrm{logit}^{-1}(\etati+d_j),
\end{equation}
where $\mathrm{logit}^{-1}(x)=(1+e^{-x})^{-1}$ and $\zeta_j=d_j$, a scalar intercept parameter.\footnote{\label{footnote:intercept}\rev{The canonical form of the Rasch model \cite<e.g.>{wright1977solving} is parameterized with an item ``difficulty'' parameter, $\delta_j=-d_j$. The parameterization in \eqref{eq:rasch} was chosen for consistency with the models in Section~\ref{sec:measurementModels}.}}
Then local independence implies that $f(\bM_{i}|\etati,\bm{d})=\prod_{j\in\mathcal{J}_i} \phi_{ij}^{M_{ij}}(1-\phi_{ij})^{1-M_{ij}}$.

We model $\etati$ as normal, conditional on covariates $\bm{x}_i$:
\begin{equation}\label{eq:etaModel}
\rev{\etati |\bx_i\sim \mathcal{N}\left(\beta_0+\boldsymbol{\beta}'\bx_i,\sigma_\eta^2\right) }
\end{equation}
with coefficient vector $\bm{\beta}$ and variance $\sigma_\eta^2$.

\rev{\label{mulitilevelGLM}
The combined model $f(\bM|\etat,\bx)f(\etat|\bx)$ implied by \eqref{eq:rasch} and \eqref{eq:etaModel} is equivalent to a standard multilevel logistic regression where $Pr(M_{ij}=1)=\mathrm{logit}^{-1}(\beta_0+\bm{x}_i'\bm{\beta}+\xi_i-d_j)$ with a student random intercept $\xi$.
Hence, the assumption that $\etati$ is conditionally normal is equivalent to the assumption of a normally distributed random intercept, which is nearly ubiquitous in multilevel modeling, and typically innocuous in that context.
That said, with Bayesian model fitting, it is a straightforward exercise to substitute an alternative conditional distribution for $\etat$; researchers suspicious of normality may estimate FLPS models using an array of alternative distributions and contrast their results.
Likewise, our broader framework can include models in which $\etat$ is discrete, categorical, and/or multivariate, which would call for alternative distributions.

The FLPS framework also allows for more elaborate measurement models, including those that include both $\etat$ and $\bx$, written in \eqref{eq:flpsMixtureT} as $f(\bM|\etat,\bx)$.
For the sake of simplicity, this manuscript will only consider models in which $\boldsymbol{x}\independent \boldsymbol{M} |\etat$, so that $f(\bM|\etat,\bx)=f(\bM|\etat)$.
This class includes models drawn from explanatory IRT \cite<c.f.>{de2013explanatory} and other generalized linear mixed models with random intercepts. }

Finally, we will assume that $Y_i$ is normally distributed conditional on $Z_i$, $\etati$, and $\bx_i$, with coefficients $\tau_0$, $\tau_1$, $\omega$, and $\bm{\gamma}$, and variance $\sigma_Y^2$:
\begin{equation}\label{eq:yModel}
\rev{Y_i|Z_i,\etati,\bx_i\sim\mathcal{N}\left(\gamma_0+\boldsymbol{\gamma}'\bx_i + \omega\etati + Z_i(\tau_0+\tau_1\etati), \sigma^2_Y\right) }.
\end{equation}
As for $\etat$, the conditional normality assumption is typical, and typically innocuous, in linear regression, but can be easily modified when it may be inappropriate.

Equation \eqref{eq:yModel} implies a linear model
for the expected treatment effect as a function
\label{text:linearEffect}
of $\etat$: $E[Y|Z=1,\etat]-E[Y|Z=0,\etat]=\tau_0+\tau_1\etat$.

In summary, the parametric FLPS models we will study here are composed of a measurement submodel $f(\bM|\etat,\bm{\zeta})$ such as \eqref{eq:rasch} or the models we will discuss in Section \ref{sec:analysis_models} and assuming local independence, and linear-normal models \eqref{eq:etaModel} with parameters $\bm{\beta}$ and $\sigma^2_\eta$ for $\etat$, conditional on $\bx$, 
and \eqref{eq:yModel}, with parameters $\bm{\gamma}$, $\omega$, $\tau_0$, $\tau_1$, and $\sigma^2_Y$ for $Y$ conditional on $\bx$, $Z$, and $\etat$. 

\section{Simulation Study}\label{sec:simulationDesign}
PS estimation, even with well-specified models, can be fraught \cite{griffin2008application,ho2022weak}; one might expect the situation to be even worse for FLPS.
To address these concerns---to establish that FLPS can be practically feasible in realistic scenarios---we conducted a Monte Carlo simulation study to investigate the operating characteristics of some parametric FLPS models.
The simulation was designed to mimic real RCT implementation data as in \citeA{datashop} and \citeA{ostrow2016assessment}.
\rev{That said, the study was designed as a proof-of-concept, testing whether under ideal circumstances---i.e. when models are regular and well-specified---FLPS can produce reliable results.
We hope to study more realistic scenarios, including model misspecification, in future work.} \label{proofOfConceptSim}
All simulation studies were carried out in R version 3.5.1 \cite{rcite} via Stan \cite{rstan}.

\subsection{Measurement Models Studied}\label{sec:measurementModels}
This study examined the feasibility of FLPS when data are generated from one of four item-response models: the Rasch model, \eqref{eq:rasch}, and
the two-parameter logistic (2PL) model \cite{birnbaum1968some} that models dichotomously scored response data, and the generalized partial credit model \cite<GPCM;>{muraki1997generalized,masters2016partial} and the graded response model \cite<GRM;>{samejima1969estimation} that models polytomously scored data.

The Rasch and 2PL models define the item response function as:

\begin{equation}\label{eq:2pl}
Pr(M_{ij}=1|\etat_i,a_j,d_j)= \mathrm{logit}^{-1}(a_j\etat_i+d_j),
\end{equation}
where $M_{ij}$ ($=m_{ij} \in \{0,1\}$) denotes the measurement outcome (i.e., response) of a subject $i$ on item $j$, $\etat_i$ ($\in \mathbb{R}$) models the subject’s latent trait level, and $a_j$ ($\in \mathbb{R}^+$) and $d_j$ ($\in \mathbb{R}$) each give the item’s slope and intercept parameters, respectively. When $a_j$ is constrained to one, the model reduces to the Rasch model. For identifying the model parameters, we constrained $a_1$ and $d_1$ 
each at one and zero.

\rev{In GPCM, measurements are ordinal, and each item $j$ has $K_j$ categories, so that $M_{ij} \in \{0, \dots , K_j \})$.
Then, the probability of student $i$ responding to score category $k$ on item $j$ is modeled as:}
\begin{equation}\label{eq:gpcm}
Pr( M_{ij} = k | \etat_i,a_j,\boldsymbol{d}_j ) = \frac{ \exp \left\{ \sum_{l=0}^k\rev{ (a_j \etat_i + d_{jl})} \right\} }{ \sum_{h=0}^{K_j} \exp \left\{ \sum_{l=0}^h \rev{(a_j \etat_i + d_{jl}) }\right\} },
\end{equation}
where \rev{$a_j\in \mathbb{R}^+$ models the trait effect on the response probability, and $d_{jl}\in \mathbb{R}$ gives the intercept of the response kernel of the category $l$ in item $j$}. Following the convention, we assume $\sum_{l=0}^0 a_j \etat_i + d_{jl} \equiv 0$. For estimating the parameters of GPCM, we constrain $a_1=1$ and $d_{11}=0$.

GRM also models the polytomously scored data but assumes strict monotonicity between the trait level and the response probability. For a subject with the trait level $\etat_i$, the probability of scoring $k$ in item $j$ is modeled by
\begin{equation}\label{eq:grm}
Pr(M_{ij} = k | \etat_i) = P_{jk}^* (\etat_i) - P_{j, k+1}^* (\etat_i),
\end{equation}
where $P_{jk}^* (\etat_i) = Pr(M_{ij} \ge k | \etat_i) = \mathrm{logit}^{-1}(a_j \etat_i + d_{jk})$. For identification, $a_1$ 
is constrained at 1 and $d_{11}$ is constrained at 0.
\subsection{Design}
Three manipulated design factors were considered to evaluate the performance of the FLPS framework: the measurement model, the sample size $N$, and the number of items $J$.
The remaining parameters were either held fixed or generated randomly.
Table~\ref{tab:tab1} summarizes the details of the factors, the condition levels, and the distributions of the randomly-generated parameters.


{\small
\begin{table}[pb!]
\centering
\caption{Simulation Design}
\label{tab:tab1}
\vspace{-0.5em}
\begin{tabular}{llll}
\toprule
Condition & Simulation factors & Values & Notation \\
\midrule
Manipulated &\hspace{3mm}Measurement model & Rasch, 2PL, GPCM, GRM  & Model \\
&\hspace{3mm}Sample size & 500, 1000, 2000  &  $N$ \\
& \hspace{3mm}Number of items & 50, 100, 200  & $J$ \\
\cmidrule(l){2-4} 
Fixed &\hspace{3mm}Number of covariates &  2 &  \\
&\hspace{3mm}Percentage of items administered &  0.6 &  \\
&\hspace{3mm}Strength of relationship &   &  \\
&\hspace{8.7mm}$\etat$ and $\yc$ &  $U$(0.1,0.3) & $\omega$ \\
&\hspace{8.7mm}$\etat$ and $\yt-\yc$ &  $U$(-0.2,-0.1) & $\tau_{1}$ \\
&\hspace{9.1mm}$Z$ and $Y$ &  $U$(0.2,0.4) & $\tau_{0}$ \\
&\hspace{9mm}$X$ and $\etat$ &  0.5 & $R^2_\eta$ \\
&\hspace{9mm}$X$ and $Y$ &  0.2 & $R^2_Y$ \\
&\hspace{3mm}Measurement model parameters &   &  \\
&\hspace{9mm}Intercept &  \makecell[l]{$N$(0,1) for binary data \\ $U$(0.5,1) for polytomous data} & $d$ \\
&\hspace{9mm}Slope &  $LogN$(0.1,1.3) & $a$ \\
\bottomrule
\end{tabular}
\begin{minipage}{\linewidth}
\vspace{0.5em}
{\footnotesize \emph{Note.} The number of items was fixed at $J=100$ when evaluating the performance under different calibration sample sizes. Similarly, the sample size was fixed at $N=1000$ when evaluating the performance under different item set sizes.}
\end{minipage}
\end{table}
}

\subsubsection{Data Generation}

Treatment group membership $Z$ and two observed covariates $\bm{x}=[X_1,X_2]'$ were generated independently: $X_1\sim N(0,1)$ and 
$X_2\sim Bernoulli(\frac{1}{2})$. 
A latent factor $\etat$ was then simulated as \eqref{eq:etaModel} with coefficients $\bm{\beta}=[-1,0.5]'$. 
We chose the residual variance $\sigma^2_\eta$ so that the covariates explain $R_\eta^2=1/2$ of the variance of $\etat$.
Next, $J$ observed implementation indicators $M_{i1},\dots,M_{iJ}$ were generated for each student $i$ in the treatment group from one of the abovementioned measurement models.
For the Rasch and 2PL models, intercept parameters $d_j$ were generated from the standard normal distribution.
The items for the GPCM and GRM were assumed to have four categories with three intercepts.
\label{intercepts}\rev{For each item $j$, intercepts $\bm{d}_j$ were generated such that $\bar{d}_j=0$ and $d_{jl+1}-d_{jl}\sim U(0.5,1)$ for $l=1,2$.
Averaged across problems, $\bar{d}_{\cdot 1}\approx -0.75$, $\bar{d}_{\cdot 2}\approx 0$, and $\bar{d}_{\cdot 3}\approx 0.75$.}
For all of the models, item slope parameters $a_j$ were generated from the log-normal distribution with a mean of 0.1 and a standard deviation of 0.3.

A random 40\% of each student's measurements were set to \textsc{na}, reflecting the average number of missing items in the RCTs we have studied.

Lastly, the outcome variable $Y$ was generated as \eqref{eq:yModel}, with $\gamma_0=0$, $\bm{\gamma}=[1,0.5]'$, $\omega$, $\tau_0$, and $\tau_1$ drawn from distributions specified in Table \ref{tab:tab1}, and $\sigma^2_Y$ chosen so that, conditional on $\etat$ and $Z$, covariates $\bm{X}$ explain $R_Y^2=0.2$ of the variance of $Y$, mimicking the RCT study of \citeA{aoas}. 

To keep the simulations at a manageable level, we conditioned the factors at a specific level as desired. For example, when examining the effect of the sample size, we fixed the number of items at $J=100$; similarly, when examining the effect of the measurement size, the sample size was fixed at $N = 1000$.
The current design resulted in 20 conditions with each condition being replicated 100 times with unique parameter sets.

\subsection{Analysis Models}\label{sec:analysis_models}

The FLPS model parameters were estimated in Stan \cite{team2016rstan} via the `rstan` package in R, which employs Hamiltonian Monte Carlo sampling. 
The models used for analysis matched the data generating models: implementation data $\bM$ were fit using the appropriate measurement model \eqref{eq:rasch}, \eqref{eq:2pl}, \eqref{eq:gpcm}, or \eqref{eq:grm}, and $\etat$ and $Y$ were modeled as \eqref{eq:etaModel} and \eqref{eq:yModel}, respectively.

We applied log-normal and standard normal prior distributions for the item slopes and intercepts, respectively.\footnote{It is based on a preliminary analysis using 40 replications comparing three priors for item slope parameters (uniform, standard normal, and log-normal) against a standard normal prior for item intercepts in a 2PL model with N = 500 and J = 50} For the structural model,  the default priors of Stan (i.e., reference distributions) were used. For the Bayesian estimation, two MCMC chains with 5000 iterations each were used to estimate the posterior distributions, with the first 2000 samples discarded in the burn-in period. The mean of the posterior distribution for each parameter was taken as its MCMC estimate.

\subsection{Evaluation}

The Gelman-Rubin convergence diagnostic (i.e., R-hat) was used to evaluate convergence, which compares the between- and within-chain estimates for model parameters \cite{gelman1992inference}. If R-hat is less than 1.1, the MCMC chains have mixed well. In this study, non-converged MCMC chains indicate those in which the R-hat of any of the model parameters was over 1.1.

To evaluate accuracy of the model parameter estimates, we examined bias and root mean squared error (RMSE):
\begin{align*}
Bias_q =& \frac{1}{R}\sum_{r=1}^{R}(\hat{\theta}_{qr}-\theta_{qr}) &
RMSE_q =& \sqrt{\frac{1}{R}\sum_{r=1}^{R}(\hat{\theta}_{qr}-\theta_{qr})^2}
\end{align*}
where $r$ ($=1,\dots, R$) indexes the replication, $\hat{\theta}_{qr}$ and $\theta_{qr}$ each indicate the estimated and generating values of the $q$th parameter of $\theta$ at replication $r$, respectively.

We also evaluated the fidelity of standard errors by examining the coverage rate of the estimated credible intervals. The credible interval was obtained as the 2.5th and 97.5th percentiles of the posterior probability distribution of each estimand. The coverage rate is the proportion of times the credible interval includes the generating parameter. 


\subsection{Results}\label{sec:simulationResults}

\subsubsection{MCMC Convergence}. The FLPS models had an overall 97.2\% convergence rate. In 56 of the 2,000 replications (20 conditions with 100 replications each), there was at least one parameter with $\hat{R}>1.1$. Among the measurement models, estimation with the 2PL showed the most frequent replications with non-convergence, (31 cases).
The FLPS with the Rasch model resulted in  100\% convergence across all replications, followed by the FLPS with the GPCM (14 cases). The GRM resulted in 11 non-converged replications. However, extreme R-hats (e.g., higher than 1.5 or 2.0) were not observed (see Section 2 in the supplementary material). Since there was no noticeable difference between the results with and without non-converged replications, all the replications were investigated including the replications with R-hats higher than 1.1.

\subsubsection{Recovery of Measurement Parameters}

Table \ref{tab:mbias} presents bias, RMSE, and coverage rates of the measurement model parameter estimates observed from the different sample-size conditions. Each entry in the table represents the average across instances of the parameter, across simulation replications (e.g. with $J=100$ items, each entry in the $a$ or $d$ columns is an average over $(J=100)\times (R=100)=10,000$ individual estimates).

The results show that the estimation overall achieved adequate accuracy and precision in the estimates. The error statistics were reasonably small, with bias close to zero in all cases. 
Moreover, as expected, RMSE tended to decrease with sample size. 
Likewise, 95\% credible intervals achieved close to their nominal levels in almost all cases---in all but one case, coverage was greater than 92\%, and in most cases it was above 94.5\%.\footnote{
The one exception was the coverage rate of 87.4\% for the $d$ parameters in the FLPS models based on GPCM measurement. We suspect this is because the intercepts were greatly influenced by the selected prior. On average, it is estimated higher or lower than the generated value set at $\pm$0.75, while the location of each of the intercept priors for the GPCM is set to 0. As a result, the first and third intercepts resulted in low coverage rates.}

\begin{table}[pt!]
\centering

\caption{Recovery of Measurement Model Parameters Under Different Sample Sizes}
\label{tab:mbias}
\vspace{-0.5em}
\begin{threeparttable}
{\small
\begin{tabular}{ccrrrrrrrrr}
\toprule
 &  & \multicolumn{3}{c}{Bias} & \multicolumn{3}{c}{RMSE} & \multicolumn{3}{c}{Coverage} \\
\cmidrule(lr){3-5} \cmidrule(lr){6-8} \cmidrule(lr){9-11}
\textit{Model} & $N_{T}$ & $a$ & $d$ & $\etat$ & $a$ & $d$ & $\etat$ & $a$ & $d$ & $\etat$ \\
\midrule
Rasch & 250 & — & 0.01 & 0.00 & — & 0.21 & 0.33 & — & 0.95 & 0.95  \\
        ~ & 500 & — & 0.01 & 0.00 & — & 0.15 & 0.33 & — & 0.96 & 0.95  \\
        ~ & 1000 & — & 0.00 & 0.00 & — & 0.11 & 0.33 & — & 0.96 & 0.95  \\
        \cmidrule{1-11}
        2PL & 250 & 0.02 & 0.00 & 0.02 & 0.32 & 0.24 & 0.42 & 0.94 & 0.95 & 0.95  \\
        ~ & 500 & 0.00 & 0.01 & 0.01 & 0.22 & 0.18 & 0.37 & 0.94 & 0.95 & 0.95  \\
        ~ & 1000 & 0.03 & 0.00 & 0.00 & 0.15 & 0.13 & 0.33 & 0.96 & 0.95 & 0.95  \\
                \cmidrule{1-11}
        GPCM & 250 & 0.04 & 0.00 & 0.01 & 0.25 & 0.42 & 0.30 & 0.93 & 0.87 & 0.95  \\
        ~ & 500 & -0.01 & 0.00 & 0.02 & 0.16 & 0.21 & 0.26 & 0.95 & 0.96 & 0.96  \\
        ~ & 1000 & 0.00 & 0.00 & 0.01 & 0.13 & 0.15 & 0.24 & 0.93 & 0.96 & 0.95  \\
                \cmidrule{1-11}
        GRM & 250 & 0.02 & -0.01 & 0.01 & 0.25 & 0.20 & 0.35 & 0.93 & 0.96 & 0.95  \\
        ~ & 500 & 0.01 & 0.00 & 0.01 & 0.18 & 0.15 & 0.31 & 0.93 & 0.96 & 0.95  \\
        ~ & 1000 & 0.00 & 0.00 & 0.01 & 0.13 & 0.10 & 0.29 & 0.92 & 0.96 & 0.95 \\
\bottomrule

\end{tabular}
}

\vspace{0.5em}
\begin{tablenotes}
    \linespread{1} \footnotesize
    \item \textit{Note}. \textit{Model}: Measurement model, $N_{T}=N/2$: Size of the treatment group. $a$: slope parameter of the item response model. $d$: intercept parameter of the item response model. $\etat$: latent trait score of the treatment group; averages are across instances of the parameter, across simulation replications; the trait estimates of the control group subjects showed average bias of .005, RMSE of 1.005, and coverage rate of .952. The number of measurement items was fixed at 100 throughout, with 40\% missing.
\end{tablenotes}
\end{threeparttable}
\vspace{-1.5em}
\end{table}

Note that results across different sample sizes are only displayed, given that outcomes under varying measurement sizes exhibited no notable differences. Further details are available in the online supplementary materials.

\subsubsection{Recovery of Structural Parameters}

Table \ref{tab:sbias} presents the bias, RMSE, and coverage rates for the structural FLPS parameters:  $\tau_0$ and $\tau_1$, the slope and intercept of causal effects as a function of $\etat$; $\bm{\beta}$, the coefficients from the $\etat$ submodel \eqref{eq:etaModel}; and $\omega$ and $\bm{\gamma}$, the coefficients of $\etat$ and covariates from the  $Y$ submodel \eqref{eq:yModel}.  

The average bias was close to zero for all structural parameters and across all conditions.
Overall, the RMSE values 
were also low---below 0.1 regardless of the simulation condition---and tended to decrease with sample size. 
RMSE ranged from 0.04 to 0.09 with $N$ = 500 and ranged from 0.02 to 0.5 with $N$ = 2000.
The ANOVA resulted in significant differences between the sample sizes with large effect sizes 
as well as medium, significant differences between conditions with different measurement models for parameters $\tau_1$ and $\bm{\beta}$ 
(effect sizes were small and non-significant for $\tau_0$, $\omega$, and $\gamma$
.

The coverage rates associated with the structural parameters were well above 0.9 across all of the conditions except for one---$\omega$ estimation under the GPCM achieved slightly lower coverage of 0.89 when $N = 2000$.



{
\begin{table}[hp]
\centering

\caption{Recovery of Structural Model Parameters Under Different Sample Sizes: Bias and RMSE}
\label{tab:sbias}
\vspace{-0.5em}
\setlength\tabcolsep{5pt}
\begin{threeparttable}
{\small
\begin{tabular}{ll rrrrr rrrrr}
\toprule
 &  & \multicolumn{5}{c}{Bias} & \multicolumn{5}{c}{RMSE} \\
\cmidrule(lr){3-7} \cmidrule(lr){8-12} 
\textit{Model} & $N$ & $\tau_0$ & $\tau_1$ & $\omega$ & $\beta$ & $\gamma$ & $\tau_0$ & $\tau_1$ & $\omega$ & $\beta$ & $\gamma$ \\ 
\midrule
Rasch & 500 & 0.00 & 0.01 & 0.00 & -0.01 & 0.01 & 0.07 & 0.07 & 0.08 & 0.07 & 0.04  \\
        ~ & 1000 & 0.00 & 0.00 & 0.00 & 0.00 & 0.00 & 0.05 & 0.05 & 0.06 & 0.05 & 0.03  \\
        ~ & 2000 & 0.00 & 0.00 & 0.00 & 0.00 & 0.00 & 0.03 & 0.03 & 0.04 & 0.04 & 0.02  \\
\cmidrule{1-12}
2PL & 500 & 0.00 & 0.00 & 0.00 & -0.03 & 0.00 & 0.07 & 0.08 & 0.09 & 0.09 & 0.04  \\
        ~ & 1000 & 0.01 & 0.00 & 0.00 & -0.01 & 0.00 & 0.05 & 0.05 & 0.06 & 0.07 & 0.03  \\
        ~ & 2000 & 0.00 & -0.01 & 0.01 & 0.00 & 0.00 & 0.04 & 0.04 & 0.05 & 0.04 & 0.02  \\
        \cmidrule{1-12}
        GPCM & 500 & 0.00 & 0.00 & 0.00 & -0.02 & 0.00 & 0.06 & 0.07 & 0.07 & 0.08 & 0.04  \\
        ~ & 1000 & 0.01 & 0.00 & 0.00 & -0.02 & 0.01 & 0.05 & 0.04 & 0.05 & 0.05 & 0.03  \\
        ~ & 2000 & 0.00 & 0.00 & 0.00 & -0.01 & 0.00 & 0.03 & 0.04 & 0.05 & 0.03 & 0.02  \\
        \cmidrule{1-12}
        GRM & 500 & 0.01 & -0.01 & 0.00 & -0.03 & -0.01 & 0.06 & 0.08 & 0.09 & 0.07 & 0.04  \\
        ~ & 1000 & 0.01 & 0.00 & 0.00 & -0.01 & 0.00 & 0.05 & 0.05 & 0.06 & 0.06 & 0.03  \\
        ~ & 2000 & 0.00 & 0.01 & 0.00 & 0.00 & 0.00 & 0.03 & 0.04 & 0.04 & 0.04 & 0.02 \\
\bottomrule
\end{tabular}
}

\vspace{0.5em}
\begin{tablenotes}
    \linespread{1} \footnotesize
    \item \textit{Note}. $N$: Sample size; the number of items was fixed at 100.
\end{tablenotes}
\end{threeparttable}
\end{table}

\begin{table}[!hp]
\centering

\caption{Structural Model Parameters: Coverage of Central 95\% Credible Intervals}
\label{tab:scoverage}
\vspace{-0.5em}
\setlength\tabcolsep{5pt}
\begin{threeparttable}
{\small
\begin{tabular}{ll rrrrr}
\toprule

\textit{Model} & $N$ & $\tau_0$ & $\tau_1$ & $\omega$ & $\beta$ & $\gamma$  \\ 
\midrule
Rasch & 500 & 0.97 & 0.94 & 0.95 & 0.95 & 0.93  \\
        ~ & 1000 & 0.92 & 0.95 & 0.96 & 0.93 & 0.97  \\
        ~ & 2000 & 0.96 & 0.97 & 0.97 & 0.96 & 0.95  \\
        \cmidrule{1-7}
        2PL & 500 & 0.97 & 0.94 & 0.94 & 0.95 & 0.97  \\
        ~ & 1000 & 0.97 & 0.97 & 0.96 & 0.96 & 0.93  \\
        ~ & 2000 & 0.95 & 0.94 & 0.94 & 0.98 & 0.97  \\
                \cmidrule{1-7}
        GPCM & 500 & 0.98 & 0.98 & 0.99 & 0.96 & 0.96  \\
        ~ & 1000 & 0.91 & 0.94 & 0.99 & 0.96 & 0.96  \\
        ~ & 2000 & 0.94 & 0.94 & 0.89 & 0.96 & 0.96  \\
                \cmidrule{1-7}
        GRM & 500 & 0.94 & 0.91 & 0.93 & 0.97 & 0.93  \\
        ~ & 1000 & 0.97 & 0.94 & 0.97 & 0.96 & 0.96  \\
        ~ & 2000 & 0.94 & 0.93 & 0.95 & 0.94 & 0.95 \\
\bottomrule
\end{tabular}
}

\vspace{0.5em}
\begin{tablenotes}
    \linespread{1} \footnotesize
    \item \textit{Note}. 
    $N$: Sample size; The number of items was fixed at 100.
\end{tablenotes}
\end{threeparttable}
\end{table}
}

\section{Implementation Dosage in a Study of Two Prealgebra Computer Programs}\label{sec:realData}

\citeA{fh2t} reported the results of a randomized efficacy study comparing four different online learning tools for 7th-grade math instruction. To illustrate FLPS, we focus on two of those conditions, which we call “Immediate” and “Delayed.” In both conditions, students worked on a series of math problems on the computer and were automatically graded. In the Immediate condition, students could request hints while working on problems, received error messages after incorrect responses, and could not proceed to the next question before submitting a correct answer. For the sake of brevity, we will refer to the hints or error messages provided by the software as ``feedback.'' Some prior work has shown such immediate feedback during practice to be beneficial (e.g. \citeNP{li2016effects}; there is extensive literature surrounding this question, e.g. \citeNP{manyclasses}, which is beyond the scope of this paper).

The “Delayed” condition resembles typical pencil-and-paper work, albeit on the computer: students attempt a series of problems and are only later given feedback on correctness. While they are working, no (automatic) hints are available.

Close to the end of the school year, students took their state’s standardized mathematics test, which we will consider the outcome of interest.

Previous analysis of this RCT has found little to no effect of immediate feedback on state test scores \cite{hintPaper}.
However, the reason for this null result is unclear: whether the two conditions are roughly equally effective, or whether students assigned to the Immediate arm of the RCT didn't receive enough feedback to make a difference, on average.
We hypothesize that if the null effect was due to low dosage, the average treatment effect of being assigned to the Immediate versus the Delayed condition should be higher for students more likely to receive feedback. 
Along those lines, we measured feedback with Rasch, 2PL, and GRM measurement models and incorporated those into larger FLPS models to estimate the extent to which treatment effects vary with students’ varying proclivities to receive tutoring.

This analysis is intended to illustrate FLPS modeling, rather than to arrive at substantive conclusions about program effectiveness.

\subsection{Data}

In the \citeA{fh2t} study, 1,141 7th-grade students were blocked within classrooms, given an online pretest consisting of ten math questions,  and individually randomized between the two conditions. We excluded students who were missing either their pretest or their outcome measurements, leaving $N$=804 students; all other variables were imputed using a random forest imputation algorithm \cite{stekhoven2012missforest}. Sample statistics for study variables---covariates, feedback, and post-test scores---are in Table \ref{tab:summaryStats}. Replication data is available by following the instructions at \url{https://osf.io/r3nf2/}.

Students from both treatment groups worked on the same set of problems within the tutor, but since our goal was to measure hints and feedback, we only modeled log data from the Immediate group, and from problems for which hints were available and responses were marked either correct or incorrect. There were 212 different problems organized into nine “problem sets.” Many problems had several parts, each of which had its own hints and was marked correct or incorrect separately; hence, we modeled each problem part---$J$=298 in total---on its own.

\begin{figure}
\centering
\includegraphics[width=0.8\textwidth]{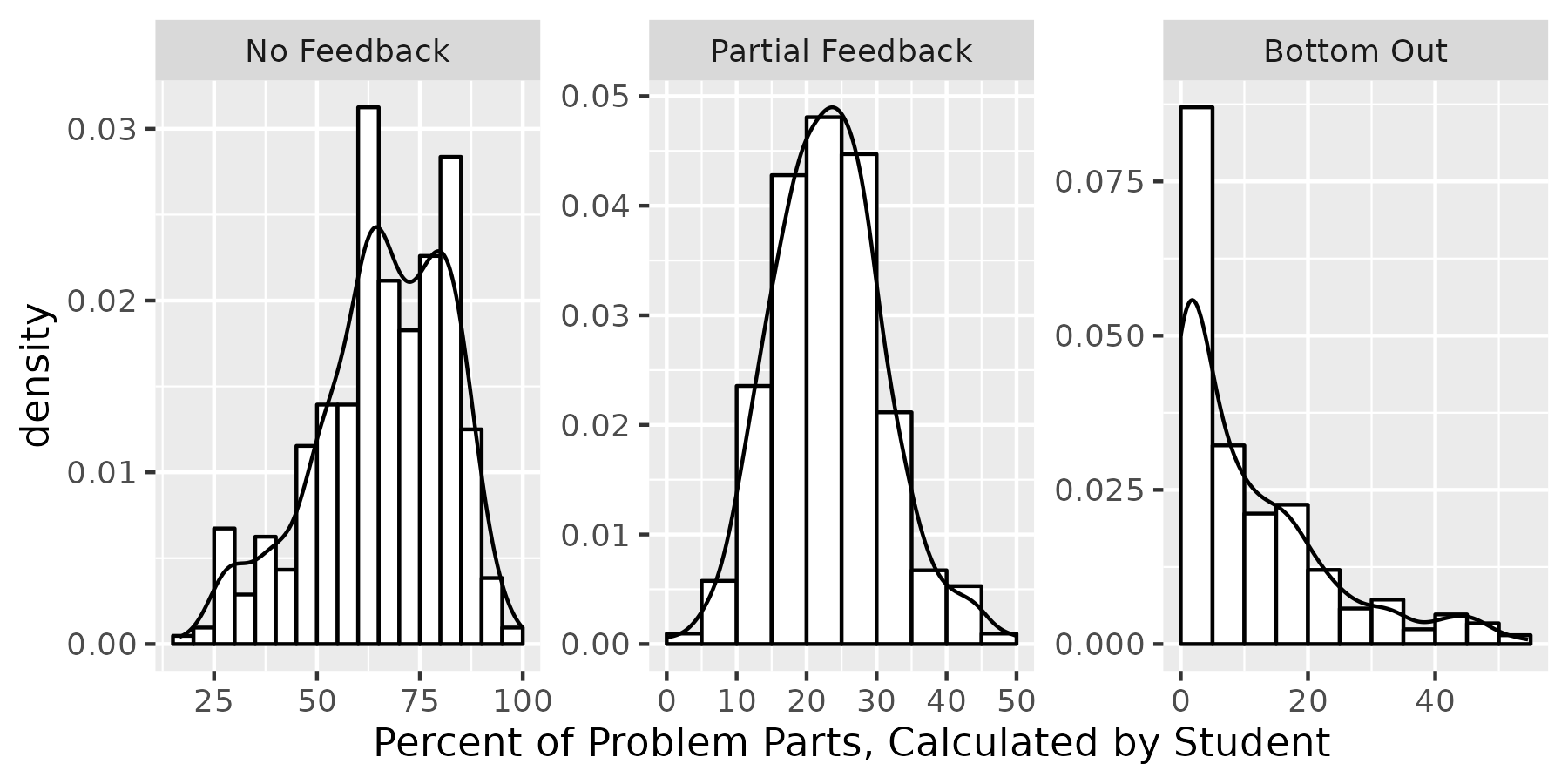}
\caption{Histograms of student-level percentages of problems on which a student received different levels of feedback.}
\label{fig:feedbackHistograms}
\end{figure}

For the Rasch and 2PL models, we dichotomized the feedback received in each worked problem part–in roughly 34\% of worked problem parts, the student received feedback by requesting at least one hint and/or committing at least one error, while in the remaining 66\% of cases, the student answered correctly on the first try without requesting a hint. This percentage varies considerably between students–-the left panel of Figure \ref{fig:feedbackHistograms} shows a histogram of the percentage of problem parts each student in the Immediate condition answered correctly on the first try. The average student received no feedback on roughly two-thirds of problems, and about half of all students in the Immediate condition received no feedback on between 60 and 80\% of problem parts. For the GRM, we also incorporated information on the amount of feedback received in each worked problem part. Depending on the problem, students could request between one and seven hints and commit up to 65 errors. To allow students to progress through the problem set, the last available “bottom out” hint contains the answer. Hence, for the GRM, we operationalized feedback received into three ordered categories: none (66\% of worked problem parts), received feedback but not the answer (22\%), and requested a bottom-out hint (12\%). The between-student variation in these percentages can be seen in the middle and rightmost panels of Figure \ref{fig:feedbackHistograms}. Almost 10\% of students in the Immediate condition requested a bottom-out hint in fewer than 5\% of the problem parts they worked.

\subsection{Models}

We fit three FLPS models to the data
in the mold of those in Sections \ref{sec:models} and \ref{sec:analysis_models}, with Rasch, 2PL, and GRM measurement submodels.
Other than the measurement submodels, the three FLPS models were the same. $\etat$ was modeled as \eqref{eq:etaModel}---normal with means linear in covariates, which included fixed teacher effects, administrative demographic and prior achievement variables, and baseline ability and non-cognitive student measures gathered during the study (a complete list of covariates is available in an online appendix). 
Outcomes $Y$ were modeled as \eqref{eq:yModel}, with the same set of covariates, as well as terms for treatment group $Z_i$, $\etat$, and their interaction.


The models were all fit in Stan, and convergence was checked by inspecting parameters’ $\hat{R}$ values and traceplots for important parameters of interest. Code in \textit{R} and \textit{Stan} is available on Github at [REDACTED].


\subsection{Results}


\begin{figure}
\centering
\input{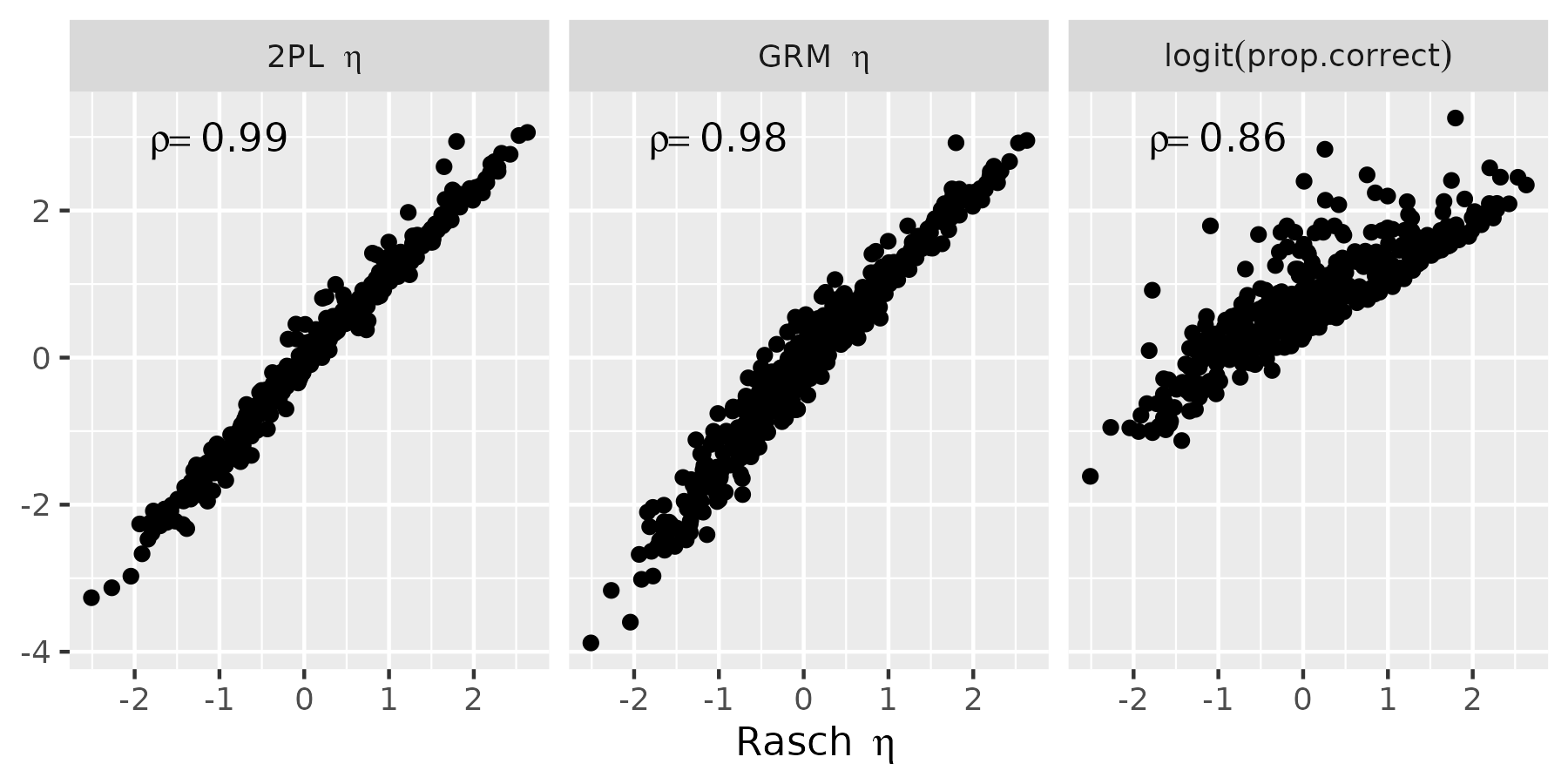}
\caption{The proportion of questions each student answered correctly on the first try and the posterior mean of $\etat$ for each student from the 2PL and GRM models plotted against the student posterior means of $\etat$ from the Rasch model}
\label{fig:rhoCompare}
\end{figure}

The latent variable $\etat$ measures differences in implementation between students---students with higher $\etat$ tended to receive more feedback. Figure \ref{fig:rhoCompare}  compares estimates (posterior means) of  $\etat$ in the Rasch model to analogous estimates from the 2PL and GRM models, respectively, as well as to the 
proportion of problems each student answered correctly on the first try,  $\hat{p}$. 
The three latent variable models agree on  $\etat$ to a remarkable extent–the Pearson correlations between estimates across Rasch, 2PL, and GRM models are all above 0.98. The correspondence to the observed $\hat{p}$ is more moderate, though also high, with a correlation of around 0.86.
The posterior standard deviations of $\etat$ between the Rasch, 2PL, and GRM models were also highly correlated, with Pearson correlations of roughly 0.99.

Table \ref{tab:usageReg} shows the results of the submodel predicting $\etat$ 
as a function of Z-scored student-level covariates. The covariates–-including teacher indicators-–were, collectively, fairly predictive, explaining 60--80\% of the variance in $\etat$. The most important predictor appears to be students’ scores on the state 5th-grade standardized math test, with students who scored higher being less likely to receive feedback. Student pretest scores showed a similar pattern, although with smaller coefficients. Students who identified as male, students in early intervention programs (EIP), students who spent less time on task during the pretest, and students who scored lower on a perceptual sensitivity may have been less likely than their peers to answer problems correctly on the first try (coefficients for these predictors were significant with p<0.05 for some models but not others, but point estimates were similar across models). 


\textit{Modeling Outcomes and Estimating Effects}
\begin{table}
\centering
\caption{Structural Parameters from outcome submodels. 
}
\label{tab:outcomeRegSmall}
\vspace{-0.5em}
\begin{threeparttable}
\begin{tabular}{lllllll}
  \hline
  & Rasch &  & 2PL &  & GRM &  \\
   \hline
$\omega$ & 0.46 * & (0.06) & 0.35 * & (0.04) & 0.32 * & (0.05) \\
  $\tau_0$ & --0.01 & (0.04) & 0.02 & (0.04) & 0.03 & (0.04) \\
  $\tau_1$ & 0.03 & (0.04) & 0.06 & (0.04) & 0.04 & (0.04) \\
   \hline
\end{tabular}
\vspace{0.5em}
\begin{tablenotes}
    \linespread{1} \footnotesize
    \item \textit{Note}. * Central 95\% credible interval excludes 0. Covariate coefficients and teacher fixed effects omitted.
\end{tablenotes}
\end{threeparttable}
\vspace{-1.5em}
\end{table}
Table \ref{tab:outcomeRegSmall} shows estimates of causal parameters $\tau_0$  and $\tau_1$ and structural parameter $\omega$ from the three FLPS models. Table \ref{tab:outcomeReg} additionally shows covariate coefficients $\boldsymbol{\gamma}$ (excluding teacher fixed effects). The outcome was standardized prior to fitting the models, so the treatment effects are in standard deviation units. The parameter $\tau_0$ can be thought of as the average treatment effect when $\etat$ 
takes its mean value–in other words, the average effect of assignment to the “Immediate” condition for the average student. The 2PL and GRM estimate small positive effects while the Rasch FLPS model estimates a small negative effect; however, the standard errors on all four estimates are sufficiently large that both small positive and negative effects are consistent with the data–in fact, all four models rule out effects for average students that are greater than 0.1 standard deviations in either direction.

Counterintuitively, all three models estimate positive $\tau_1$: that the effect of assignment to the “Immediate” condition is higher, on average, for students who had a greater propensity to answer problems correctly on the first try, and hence to receive less feedback. However, the data are also consistent with values of $\tau_1$ that are slightly negative or null–the posterior probability that $\tau_1>0$ ranges from 0.81 (Rasch) to 0.96 (2PL).

A positive value of $\tau_1$ would not necessarily imply that receiving feedback decreases the effect of the treatment assignment; in the principal stratification framework, $\etat$ is taken as a student baseline characteristic rather than a manipulable behavior. Students with high values of $\etat$ may differ from their low-$\etat$ peers in other ways that may lead to positive effects of immediate feedback. For instance, students who answer problems correctly with greater frequency may be more conscientious and may benefit more from the feedback they get. 

\begin{figure}
\centering
\input{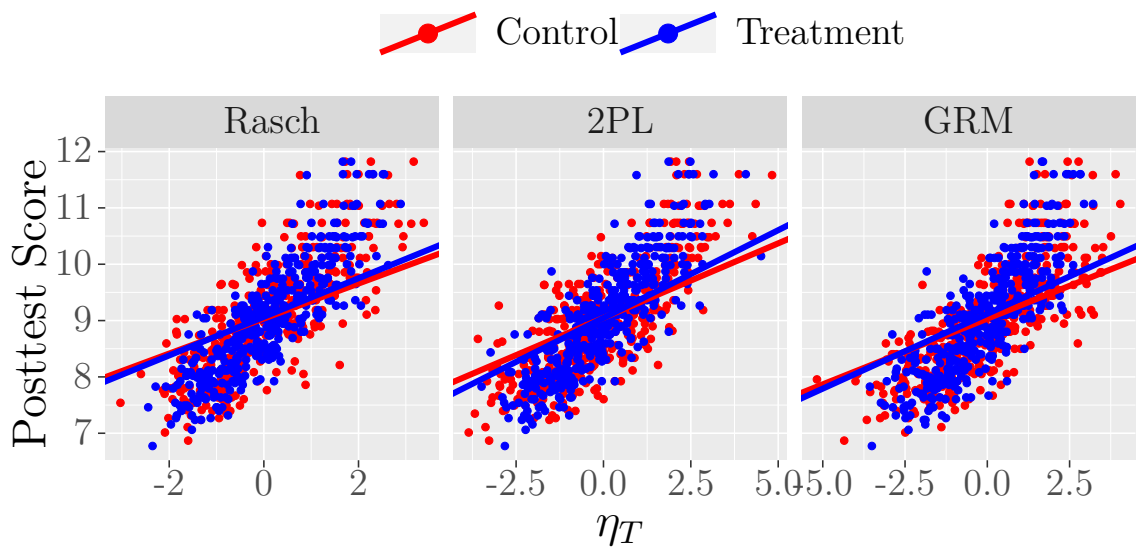}
\caption{One random draw of $\etat$ (estimated for the Immediate group, imputed for the Delayed group) plotted against observed posttest scores. The distance between the two trend-lines at a value of $\etat$ is the principal effect.}
\label{fig:etaYmodel}
\end{figure}

The moderator-like role of $\etat$ in principal stratification is illustrated in Figure \ref{fig:etaYmodel}, which plots a random posterior draw of $\etat$, as measured or imputed in each FLPS model, against posttest scores $Y$. 
In all three models, for both treatment arms, $\etat$ is positively correlated with posttest scores. 
However, the slopes are slightly different between the two treatment arms–--that is, there is an interaction between $\etat$ and the treatment indicator $Z$. 

We hypothesized that if the null average effect of immediate feedback on state test scores was due to low average dosage,
 we would estimate  $\tau_1<0$ in FLPS models since $\etat$ is higher for students who received less feedback. Instead, we estimated $\tau_1$ to be close to zero; if anything, $\tau_1$ was more likely to be positive than negative.
 These results suggest that low average dosage cannot explain the null finding.

\label{robust1}
In fitting FLPS models, it made little difference what measurement model to use---estimates of $\etat$, $\tau_0$, and $\tau_1$ were highly similar between the three models.
This comforting regularity suggests a robustness to misspecification of the measurement model; future research may determine whether this reflects an underlying property of FLPS models or was merely fortuitous.
In any event, the results of this illustration underscore the conclusions in Section \ref{sec:simulationResults}, that FLPS models are feasible in real-life data analyses.


\section{Discussion}\label{sec:discussion}

Measurement of program implementation and implementation fidelity in an RCT can be complex and nuanced, combining different types of data and different data sources \cite<e.g.>{goodson2014measuring,schoenwald2011toward,carroll2007conceptual,vanacore_ottmar_liu_sales_2023}.
The rich log data produced by technology-based interventions such as EdTech demands even greater care in implementation modeling.
This paper evaluated FLPS, a new approach to incorporating these high-quality measurements of implementation into the causal modeling process, allowing researchers to estimate the moderating role of implementation in an intervention's effectiveness.
Given an implementation model, FLPS attempts to use baseline covariates to impute potential implementation, $\etat$, for subjects assigned to the control condition, and then estimate an interaction between $\etat$ and treatment assignment in a causal model.
Bayesian model fitting, which considers the implementation, imputation, and causal models as interconnected facets of a larger model, ensures that uncertainty is properly propagated.

Or, at least, it should.
It can be difficult to tell if complex models such as the FLPS truly estimates what we want them to estimate.
This is especially true in the PS context, where there is cause for concern that even well-specified models may yield severely biased estimates \cite{griffin2008application,ho2022weak}.
These concerns motivated our extensive simulation study.


Our simulation study showed that the FLPS worked properly across the simulation conditions with different IRT models as measurement models. 
Specifically, all model parameter estimates---and in particular,  principal effect, which is the primary concern of principal stratification---were estimated with low bias and adequate accuracy across the conditions. The most influential factor in determining RMSE was sample size---increasing sample size leads to more accurate estimation---followed by the type of measurement model. 
Our estimated credible intervals also performed well, with
coverage rates over 0.9 in all conditions except for one. 

We also illustrated the use of the FLPS in analyzing data from an RCT comparing an EdTech application that provided immediate feedback to struggling students to one that did not.
There appeared to be little to no effect of immediate feedback on students' state test scores---our FLPS results showed that this null result was probably not due to low dosage (i.e. too little feedback), but likely reflected ineffectiveness for other reasons.
All three measurement models we considered gave strikingly similar estimates, suggesting that, at least in this case, the FLPS was robust to minor misspecification of the measurement model.

The results of this study suggest that FLPS models can allow researchers to gain deeper and more nuanced insights into the relationship between the effectiveness of educational interventions and various aspects of their implementation---in particular when big implementation data is available, such as in evaluations of computer-based interventions.
Some recently-published applications have begun to bear this out:
\citeA{gamers} used an FLPS model to show that providing immediate feedback  and computer-based tutoring in educational technology may be counterproductive for students who are abusing those resources, and \citeA{replay} used FLPS to show that
an educational computer program that incorporated gamification was more beneficial to students who were more likely to redo problems after suboptimal performance.

The success of those analyses and of the models we considered here suggests even more exciting possibilities.
First, the current study only considered IRT models, which apply to categorical indicators. Other types of log data vary in measurement, such as continuous, longitudinal, and more complex data structures. Future work will extend the model to confirmatory factor analysis or latent growth modeling.
Indeed, many different types of latent variable models are available for FLPS. The current study only considered the continuous latent variable based on the IRT models. Since discrete latent variables are possible options for the principal effects, we hope that future work will also incorporate mixture models into FLPS such as latent class analysis modeling or factor mixture modeling.

Furthermore, this study was only conducted in terms of a single latent variable. There are more complicated relations within log-data, which multidimensional structures may underlie. It will be worth exploring model formulations for a variety of multidimensional factor models to be included in the FLPS framework in future research.

\label{assumptionsDiscussion}
\rev{The estimators we presented here depended on several strong parametric assumptions, such as \eqref{eq:independenceY}.
Due to the inherent difficulty of reasoning about unobserved potential measurements, such as $M^T$, $M^C$, or $\etat$, this is fairly typical.
Most PS methods rely on complex parametric models \cite<e.g.>[]{roy2007principal,zhang2009likelihood,li2010bayesian,page2012principal} and/or strong, untestable structural assumptions such as the exclusion restriction and monotonicity \cite{AIR} or principal ignorability \cite{ding2017principal,feller2017principal}.
Assumptions such as \eqref{eq:independenceY} about model adequacy are often testable, in principle, unlike the structural assumptions of, say, \citeA{AIR}.
Moreover, analysts suspicious of a model are free to specify an alternative.
That said, it is often unclear how best to test a model specification, what sample size would be required for an adequately powered test, or what alternative models to attempt.

\label{futureResearch}
These challenges call for future research in at least three directions.
First, they call for the development of FLPS estimators that are less reliant on parametric model specifications---the identification results from \citeA<e.g.>[]{jiangDing2021} and the moment estimator of \citeA{sales2022geepers}, both developed for classical PS, may be helpful starting points.
A second direction is the development and validation of model-checking procedures, perhaps building on those suggested in \citeA{aoas}.
Finally, 
extending the simulation study of Section \ref{sec:simulationDesign} to cases with model misspecification could clarify the sensitivity or robustness of FLPS estimation to violations of its assumptions.

\label{reviewersIdea}
More broadly, there is an increasing need for methodological research on any tool to help behavioral researchers make better sense of complex implementation data.
For instance, one anonymous reviewer suggested that researchers could model $\Mt$ or $\etat$ as treatments in and of themselves, instead of modeling them as latent moderators of treatment effects, as in PS and FLPS.
This approach bears a lot in common with \citeA{VanderWeeleHernan+2013+1+20}, which considers multiple versions of a treatment, and with \citeA{sales2021student}, which contrasts an FLPS model with analyses of $\Mt$ as an observational study embedded within an experiment or as a causal mediator.
As described there, each approach has its advantages and disadvantages, answers different research questions, and can play a valuable role in analyses of implementation data. }

Clearly, subjects who would implement an intervention differently will experience different treatment effects; the role of FLPS is to use the complex implementation data already being gathered to measure those differences.
While deep questions remain about the robustness and scope of FLPS estimation, this paper demonstrated that FLPS is feasible, and can be a valuable tool in behavioral effectiveness research.

\vspace{5mm}


%
%
\setcounter{secnumdepth}{0}
\setstretch{1}
\bibliography{flps}

\newpage
\appendix

\begin{center}
    {\large \textbf{
    Online Supplement to\\ ``Fully Latent Principal Stratification With Measurement Models''\\
    Details and Results from Simulation and Applied Studies}
    }
\end{center}

\section{Details and Results from the Simulation Study}

\subsection{Pilot Study}

\setlength{\LTpost}{0mm}
\begin{longtable}{ccccccccc}
\caption{Summary of Pilot Study: measurement parts} \\
\toprule
 &  \multicolumn{4}{c}{Bias} & \multicolumn{4}{c}{RMSE} \\
\cmidrule(lr){2-5} \cmidrule(lr){6-9}
Prior & a & d & $\eta_T$ & $\eta_C$ & a & d & $\eta_T$ & $\eta_C$ \\
\midrule
LogN & 0.069 & -0.017 & 0.025 & 0.017 & 0.340 & 0.232 & 0.473 & 1.011  \\
Normal  & -0.193 & 0.014 & 0.052 & 0.064 & 0.282 & 0.234 & 0.547 & 1.039 \\
Unif  & 4.600 & -0.018 & -0.191 & -0.191 & 4.872 & 0.242 & 1.210 & 1.326 \\
\bottomrule
\end{longtable}
\vspace{0.1in}
\begin{center}
\begin{minipage} {5in} 
\hspace{0in}\textbf{\emph{Note.}} Model: 2PL; N: 500; J: 50; a: item slope; d: item intercept; $\eta_T$: latent factor scores for the treatment group; $\eta_C$: latent factor scores for the control group\\
\end{minipage}
\end{center}

\setlength{\LTpost}{0mm}
\begin{longtable}{ccccccc}
\caption{Summary of Pilot Study: structural parts}\\
\toprule
 & \multicolumn{3}{c}{Bias} & \multicolumn{3}{c}{RMSE}  \\
\cmidrule(lr){2-4} \cmidrule(lr){5-7}
Prior  & $\tau_0$ & $\tau_1$ & $\omega$ & $\tau_0$ & $\tau_1$ & $\omega$ \\
\midrule
LogN  & -0.013 &  0.004 & -0.003 & 0.074 & 0.069 & 0.088 \\
Normal  & -0.001 &  0.042 & -0.027 & 0.063 & 0.075 & 0.074\\
Unif  & -0.006 & -0.754 &  0.822 & 0.075 & 0.826 & 0.910 \\
\bottomrule
\end{longtable}
\vspace{0in}
\begin{center}
\begin{minipage}{0.6\linewidth}
\textbf{\emph{Note.}} Note. Model: 2PL; N: 500; J: 50; $\tau_0$: the difference between the treatment and the control group; $\tau_1$: principal effect; $\omega$: the latent factor effect on the outcome
\end{minipage}
\end{center}

\pagebreak

\subsection{R-hat for all parameters across the simulation conditions}

\setlength{\LTpost}{0mm}

\begin{longtable}{ccccccccccc}
\caption{R-hat for all parmaeters across the simulation conditions}\\
\toprule
N & J & PN & \multicolumn{4}{c}{Mean} & \multicolumn{4}{c}{Max} \\
\cmidrule(lr){4-7} \cmidrule(lr){8-11}
&  &  & Rasch & 2PL & GPCM & GRM & Rasch & 2PL & GPCM & GRM \\

\midrule
500 & 100 & $\omega$ & 1.003 & 1.014 & 1.006 & 1.007 & 1.017 & 1.062 & 1.027 & 1.054 \\
500 & 100 & $\tau_0$ & 1.000 & 1.000 & 1.000 & 1.000 & 1.003 & 1.004 & 1.002 & 1.002 \\
500 & 100 & $\tau_1$ & 1.002 & 1.018 & 1.007 & 1.010 & 1.014 & 1.119 & 1.046 & 1.065 \\
500 & 100 & $\beta$ & 1.002 & 1.037 & 1.019 & 1.024 & 1.012 & 1.114 & 1.116 & 1.100  \\
500 & 100 & $\gamma$ & 1.001 & 1.001 & 1.000 & 1.000 & 1.005 & 1.013 & 1.005 & 1.006 \\
500 & 100 & $\eta_C$ & 1.000 & 1.002 & 1.001 & 1.001 & 1.001 & 1.007 & 1.005 & 1.004 \\
500 & 100 & $\eta_T$ & 1.000 & 1.015 & 1.012 & 1.011 & 1.002 & 1.048 & 1.074 & 1.043 \\
500 & 100 & $a$ & — & 1.027 & 1.016 & 1.019 & — & 1.090 & 1.124 & 1.098 \\
500 & 100 & d & 1.001 & 1.001 & 1.000 & 1.000 & 1.004 & 1.002 & 1.000 & 1.001        \\
\hline
1000 & 50 & $\omega$ & 1.002 & 1.013 & 1.006 & 1.007 & 1.017 & 1.076 & 1.082 & 1.042 \\
1000 & 50 & $\tau_0$ & 1.000 & 1.000 & 1.000 & 1.000 & 1.002 & 1.002 & 1.002 & 1.001 \\
1000 & 50 & $\tau_1$ & 1.001 & 1.018 & 1.008 & 1.009 & 1.011 & 1.099 & 1.073 & 1.040 \\
1000 & 50 & $\beta$ & 1.001 & 1.033 & 1.021 & 1.020 & 1.007 & 1.153 & 1.128 & 1.076  \\
1000 & 50 & $\gamma$ & 1.001 & 1.001 & 1.000 & 1.001 & 1.007 & 1.004 & 1.003 & 1.002 \\
1000 & 50 & $\eta_C$ & 1.000 & 1.001 & 1.000 & 1.000 & 1.000 & 1.007 & 1.004 & 1.001 \\
1000 & 50 & $\eta_T$ & 1.000 & 1.006 & 1.006 & 1.004 & 1.000 & 1.039 & 1.032 & 1.015 \\
1000 & 50 & $a$ & — & 1.024 & 1.017 & 1.016 & — & 1.144 & 1.097 & 1.060 \\
1000 & 50 & d & 1.000 & 1.001 & 1.000 & 1.000 & 1.003 & 1.003 & 1.002 & 1.002         \\
\hline
1000 & 100 & $\omega$ & 1.002 & 1.015 & 1.008 & 1.006 & 1.010 & 1.071 & 1.088 & 1.025 \\
1000 & 100 & $\tau_0$ & 1.000 & 1.000 & 1.000 & 1.000 & 1.002 & 1.002 & 1.002 & 1.001 \\
1000 & 100 & $\tau_1$ & 1.001 & 1.018 & 1.010 & 1.009 & 1.008 & 1.072 & 1.144 & 1.039 \\
1000 & 100 & $\beta$ & 1.002 & 1.037 & 1.030 & 1.023 & 1.010 & 1.139 & 1.210 & 1.124  \\
1000 & 100 & $\gamma$ & 1.001 & 1.001 & 1.000 & 1.001 & 1.004 & 1.004 & 1.003 & 1.004 \\
1000 & 100 & $\eta_C$ & 1.000 & 1.001 & 1.000 & 1.000 & 1.000 & 1.004 & 1.004 & 1.003 \\
1000 & 100 & $\eta_T$ & 1.000 & 1.010 & 1.012 & 1.007 & 1.001 & 1.040 & 1.078 & 1.039 \\
1000 & 100 & $a$ & — & 1.028 & 1.024 & 1.018 & — & 1.099 & 1.157 & 1.110  \\
1000 & 100 & d & 1.001 & 1.001 & 1.000 & 1.000 & 1.005 & 1.004 & 1.001 & 1.001        \\
\hline
1000 & 200 & $\omega$ & 1.002 & 1.014 & 1.010 & 1.008 & 1.011 & 1.112 & 1.048 & 1.040 \\
1000 & 200 & $\tau_0$ & 1.000 & 1.000 & 1.000 & 1.000 & 1.003 & 1.003 & 1.001 & 1.001 \\
1000 & 200 & $\tau_1$ & 1.001 & 1.019 & 1.012 & 1.011 & 1.009 & 1.101 & 1.064 & 1.039 \\
1000 & 200 & $\beta$ & 1.002 & 1.046 & 1.044 & 1.034 & 1.013 & 1.202 & 1.205 & 1.129   \\
1000 & 200 & $\gamma$ & 1.001 & 1.000 & 1.000 & 1.000 & 1.004 & 1.003 & 1.003 & 1.002  \\
1000 & 200 & $\eta_C$ & 1.000 & 1.001 & 1.001 & 1.001 & 1.000 & 1.005 & 1.003 & 1.002  \\
1000 & 200 & $\eta_T$ & 1.000 & 1.018 & 1.026 & 1.015 & 1.003 & 1.071 & 1.098 & 1.066  \\
1000 & 200 & $a$ & — & 1.032 & 1.033 & 1.027 & — & 1.129 & 1.119 & 1.114 \\
1000 & 200 & d & 1.001 & 1.001 & 1.000 & 1.000 & 1.006 & 1.004 & 1.001 & 1.001        \\
\hline
2000 & 100 & $\omega$ & 1.001 & 1.014 & 1.008 & 1.007 & 1.007 & 1.064 & 1.045 & 1.034 \\
2000 & 100 & $\tau_0$ & 1.000 & 1.000 & 1.000 & 1.000 & 1.002 & 1.002 & 1.001 & 1.001 \\
2000 & 100 & $\tau_1$ & 1.001 & 1.020 & 1.011 & 1.009 & 1.005 & 1.123 & 1.040 & 1.031 \\
2000 & 100 & $\beta$ & 1.002 & 1.041 & 1.030 & 1.030 & 1.010 & 1.190 & 1.103 & 1.110   \\
2000 & 100 & $\gamma$ & 1.001 & 1.000 & 1.000 & 1.000 & 1.004 & 1.003 & 1.004 & 1.002  \\
2000 & 100 & $\eta_C$ & 1.000 & 1.001 & 1.000 & 1.000 & 1.000 & 1.003 & 1.001 & 1.001  \\
2000 & 100 & $\eta_T$ & 1.000 & 1.007 & 1.007 & 1.005 & 1.001 & 1.036 & 1.024 & 1.020  \\
2000 & 100 & $a$ & — & 1.030 & 1.023 & 1.024 & — & 1.150 & 1.072 & 1.091 \\
2000 & 100 & d & 1.001 & 1.001 & 1.001 & 1.001 & 1.007 & 1.004 & 1.003 & 1.004 \\
\bottomrule
\end{longtable}
\vspace{0.1in}
\begin{center}
\begin{minipage}{\linewidth}
\hspace{0.1in}\emph{Note.} N: measurement model; J: Number of items; PN: parameter name\\
\end{minipage}
\end{center}

\subsection{Freq of non-convergence MCMC results by parameters}

\setlength{\LTpost}{0mm}
\begin{longtable}{ccccccccc}
\caption{Counts of non-convergence MCMC results}\\
\toprule
 &  &  & b11 & bu & lambda & NA & a11 & tau \\
\cmidrule(lr){4-4} \cmidrule(lr){5-5} \cmidrule(lr){6-6} \cmidrule(lr){7-7} \cmidrule(lr){8-8} \cmidrule(lr){9-9}
MM & N & J & Freq & Freq & Freq & Freq & Freq & Freq \\
\midrule
2PL & 500 & 100 & 1 & 3 & 3 & 0 & 0 & 0 \\
2PL & 1000 & 50 & 0 & 3 & 2 & 1 & 0 & 0 \\
2PL & 1000 & 100 & 0 & 6 & 6 & 0 & 0 & 0 \\
2PL & 1000 & 200 & 1 & 10 & 9 & 8 & 1 & 0 \\
2PL & 2000 & 100 & 2 & 6 & 5 & 0 & 0 & 0 \\
GPCM & 500 & 100 & 0 & 1 & 1 & 1 & 0 & 0 \\
GPCM & 1000 & 50 & 0 & 1 & 1 & 0 & 0 & 0 \\
GPCM & 1000 & 100 & 1 & 1 & 1 & 1 & 0 & 1 \\
GPCM & 1000 & 200 & 0 & 8 & 8 & 9 & 0 & 0 \\
GPCM & 2000 & 100 & 0 & 1 & 0 & 0 & 0 & 0 \\
GRM & 500 & 100 & 0 & 0 & 1 & 0 & 0 & 0 \\
GRM & 1000 & 100 & 0 & 2 & 2 & 0 & 0 & 0 \\
GRM & 1000 & 200 & 0 & 4 & 5 & 3 & 0 & 0 \\
GRM & 2000 & 100 & 0 & 3 & 2 & 0 & 0 & 0 \\
\bottomrule
\end{longtable}
\vspace{0.1in}
\begin{center}
\begin{minipage}{\linewidth}
\hspace{0.8in}\emph{Note.} N: sample size; J: number of item\\
\end{minipage}
\end{center}

\subsection{Simulation results under varying measurement sizes}

\subsubsection{Recovery of Measurement Model Parameters}

Table \ref{tab:mbias_online} presents evaluation statistics of the measurement model estimates under the different measurement-size conditions. As expected, longer assessments entailed more precise trait recovery. As the number of items increased from 30 to 60 and 120, the biasedness of the trait estimates stayed around zero (0.08 on average) while RMSE decreased from 0.402 to 0.317 and 0.250 on average. The coverage rate of the interval estimates was constantly kept at the nominal level, averaging 0.951 rate.

\begin{table}[h]
\centering
\caption{Recovery of Measurement Model Parameters Under Different Measurement Sizes}
\label{tab:mbias_online}
\vspace{-0.5em}
\begin{threeparttable}
{\small
\begin{tabular}{ccrrrrrrrrr}
\toprule
 &  & \multicolumn{3}{c}{Bias} & \multicolumn{3}{c}{RMSE} & \multicolumn{3}{c}{Coverage} \\
\cmidrule(lr){3-5} \cmidrule(lr){6-8} \cmidrule(lr){9-11}
\textit{Model} & $J$ & $a$ & $b$ & $\eta_T$ & $a$ & $b$ & $\eta_T$ & $a$ & $b$ & $\eta_T$ \\
\midrule
Rasch & 50 & — &  0.006 & -0.005 & — & 0.154 & 0.436 & — & 0.950 & 0.951 \\
 & 100 & — &  0.005 & -0.004 & — & 0.152 & 0.330 & — & 0.956 & 0.949 \\
 & 200 & — & -0.003 &  0.002 & — & 0.151 & 0.243 & — & 0.948 & 0.950 \\
\cmidrule{1-11}
2PL & 50 &  0.032 &  0.001 &  0.007 & 0.241 & 0.176 & 0.458 & 0.929 & 0.956 & 0.949 \\
 & 100 &  0.001 &  0.005 &  0.010 & 0.220 & 0.178 & 0.369 & 0.937 & 0.946 & 0.951 \\
 & 200 &  0.009 & -0.005 &  0.018 & 0.222 & 0.169 & 0.299 & 0.936 & 0.953 & 0.948 \\
\cmidrule{1-11}
GPCM & 50 & -0.004 & -0.004 &  0.007 & 0.166 & 0.212 & 0.320 & 0.934 & 0.956 & 0.955 \\
 & 100 & -0.007 &  0.003 &  0.015 & 0.164 & 0.209 & 0.259 & 0.945 & 0.957 & 0.955 \\
 & 200 & -0.015 &  0.000 &  0.013 & 0.160 & 0.241 & 0.214 & 0.944 & 0.936 & 0.950 \\
\cmidrule{1-11}
GRM & 50 &  0.000 & -0.003 &  0.009 & 0.183 & 0.147 & 0.393 & 0.921 & 0.959 & 0.953 \\
 & 100 &  0.005 & -0.001 &  0.011 & 0.180 & 0.145 & 0.309 & 0.927 & 0.959 & 0.948 \\
 & 200 & -0.004 & -0.005 &  0.007 & 0.166 & 0.144 & 0.244 & 0.944 & 0.957 & 0.953 \\
\bottomrule
\end{tabular}
}
\vspace{0.5em}
\begin{tablenotes}
    \linespread{1} \footnotesize
    \item \emph{Note.} Measurement model, $J$: Number of items; With 40\% missing, 60\% of $J$ were actually used for calibration, $a$: Slope parameter of the item response model. $d$: Intercept parameter of the item response model. $\eta_{T}$: Latent trait score of the treatment group; The sample size of the treatment group ($N_T$) was fixed at 500. The trait estimates of the control group subjects showed average bias of .010, RMSE of 1.005, and coverage rate of .953.
\end{tablenotes}
\end{threeparttable}
\vspace{-1.5em}
\end{table}

\subsubsection{Recovery of Structural Model Parameters}

Tables \ref{tab:sbias_online} present the results related to the structural paths in terms of bias, RMSE, and coverage rates. The parameters examined include: the direct effect of $Z$ to $Y$ ($\tau_0$), the principal effect ($\tau_1$), the effect of the latent trait on $Y$ ($\omega$), the effect of the covariates on the latent trait ($\beta$), and the effect of the covariates on Y ($\gamma$). The results presented in Table \ref{tab:sbias_online} illustrate the impact of the number of items on bias, RMSE, and coverage rates with the sample size fixed at 1000. Again, the RMSE values associated with all of the structural model parameters were close to zero, falling below 0.1, regardless of the simulation condition. In contrast to the impact of sample size, the RMSE values associated with the structural parameters showed no significant or meaningful differences as the sample size is fixed at 1000 [$F(2, 8) = 0.266$, $p > 0.05$, $\eta^2 = 0.003$ for $\tau_0$; $F(2, 8) = 2.055$, $p > 0.05$, $\eta^2 = 0.013$ for $\tau_1$; $F(2, 8) = 2.054$, $p > 0.05$, $\eta^2 = 0.032$ for $\omega$; $F(2, 8) = 2.171$, $p > 0.05$, $\eta^2 = 0.022$ for $\beta$; and $F(2, 8) = 2.704$, $p > 0.05$, $\eta^2 = 0.007$ for $\gamma$]. The coverage rates associated with the structural parameters were well above 0.9 across all of the conditions.

{
\begin{landscape}
\begin{table}[h]
\centering
\caption{Recovery of Structural Model Parameters Under Different Measurement Sizes}
\label{tab:sbias_online}
\vspace{-0.5em}
\setlength\tabcolsep{5pt}
\begin{threeparttable}
\begin{tabular}{llrrrrrrrrrrrrrrr}
\toprule
 &  & \multicolumn{5}{c}{Bias} & \multicolumn{5}{c}{RMSE} & \multicolumn{5}{c}{Coverage} \\
\cmidrule(lr){3-7} \cmidrule(lr){8-12} \cmidrule(lr){13-17}
\textit{Model} & $J$ & $\tau_0$ & $\tau_1$ & $\omega$ & $\beta$ & $\gamma$ & $\tau_0$ & $\tau_1$ & $\omega$ & $\beta$ & $\gamma$ & $\tau_0$ & $\tau_1$ & $\omega$ & $\beta$ & $\gamma$ \\
\midrule
Rasch & 50 &  0.000 & -0.004 &  0.004 & -0.005 &  0.001 & 0.045 & 0.044 & 0.064 & 0.051 & 0.026 & 0.960 & 0.970 & 0.920 & 0.960 & 0.925 \\
      & 100 &  0.001 & -0.002 & -0.001 & -0.002 & -0.001 & 0.046 & 0.048 & 0.055 & 0.054 & 0.028 & 0.920 & 0.950 & 0.960 & 0.930 & 0.965 \\
      & 200 & -0.001 & -0.004 &  0.008 &  0.007 & -0.001 & 0.042 & 0.044 & 0.057 & 0.049 & 0.030 & 0.970 & 0.950 & 0.930 & 0.935 & 0.935 \\
\cmidrule{1-17}
2PL & 50 &  0.010 & -0.015 &  0.023 & -0.002 &  0.000 & 0.047 & 0.059 & 0.071 & 0.057 & 0.028 & 0.970 & 0.930 & 0.940 & 0.970 & 0.960 \\
    & 100 &  0.008 & -0.002 &  0.001 & -0.009 &  0.003 & 0.046 & 0.051 & 0.060 & 0.065 & 0.028 & 0.970 & 0.970 & 0.960 & 0.955 & 0.925 \\
    & 200 & -0.002 &  0.001 & -0.001 & -0.007 &  0.000 & 0.043 & 0.055 & 0.065 & 0.054 & 0.027 & 0.980 & 0.910 & 0.950 & 0.935 & 0.945 \\
\cmidrule{1-17}
GPCM & 50 & -0.004 &  0.009 & -0.014 & -0.018 & -0.003 & 0.051 & 0.056 & 0.065 & 0.055 & 0.028 & 0.930 & 0.930 & 0.910 & 0.970 & 0.935 \\
     & 100 &  0.005 &  0.002 &  0.003 & -0.015 &  0.005 & 0.052 & 0.044 & 0.051 & 0.048 & 0.028 & 0.910 & 0.940 & 0.990 & 0.960 & 0.955 \\
     & 200 &  0.002 &  0.002 & -0.003 & -0.015 &  0.004 & 0.049 & 0.052 & 0.064 & 0.048 & 0.030 & 0.950 & 0.920 & 0.910 & 0.960 & 0.945 \\
\cmidrule{1-17}
GRM & 50 &  0.002 &  0.008 & -0.006 & -0.009 &  0.003 & 0.047 & 0.049 & 0.054 & 0.058 & 0.026 & 0.940 & 0.950 & 0.950 & 0.955 & 0.975 \\
    & 100 &  0.007 &  0.002 & -0.002 & -0.005 &  0.001 & 0.048 & 0.046 & 0.057 & 0.057 & 0.025 & 0.970 & 0.940 & 0.970 & 0.955 & 0.960 \\
    & 200 & -0.001 &  0.004 & -0.001 & -0.017 &  0.003 & 0.051 & 0.051 & 0.059 & 0.048 & 0.028 & 0.940 & 0.970 & 0.960 & 0.960 & 0.955 \\
\bottomrule
\end{tabular}

\vspace{0.5em}
\begin{tablenotes}
    \linespread{1} \footnotesize
    \item \emph{Note.} \textit{Model}: Measurement model; $J$: Number of items; The sample size was fixed at 1000.
\end{tablenotes}
\end{threeparttable}
\end{table}

\end{landscape}
}

\pagebreak

\section{Details and Results from the Application}

\subsection{Data Description and Regression Tables}

\begin{table}[pt!]
\centering
\caption{Summary Statistics of Study Variables}
\label{tab:summaryStats}
\vspace{-0.5em}
{\small
\begin{tabular}{llll}
\toprule
  & Delayed ($N=388$) & Immediate ($N=416$) & Overall ($N=804$)\\
\midrule
\addlinespace[0.3em]
\multicolumn{4}{l}{\textbf{Pretest}}\\
\hspace{1em}Mean (SD) & 4.76 (2.67) & 4.91 (2.64) & 4.84 (2.65)\\
\addlinespace[0.3em]
\multicolumn{4}{l}{\textbf{5th Grade State Test}}\\
\hspace{1em}Mean (SD) & 581 (58.8) & 580 (59.1) & 581 (58.9)\\
\addlinespace[0.3em]
\multicolumn{4}{l}{\textbf{Sex}}\\
\hspace{1em}Female & 176 (45.4\%) & 204 (49.0\%) & 380 (47.3\%)\\
\hspace{1em}Male & 212 (54.6\%) & 212 (51.0\%) & 424 (52.7\%)\\
\addlinespace[0.3em]
\multicolumn{4}{l}{\textbf{Race/Ethnicity}}\\
\hspace{1em}White & 186 (47.9\%) & 226 (54.3\%) & 412 (51.2\%)\\
\hspace{1em}Hispanic/Latino & 57 (14.7\%) & 51 (12.3\%) & 108 (13.4\%)\\
\hspace{1em}Asian & 109 (28.1\%) & 109 (26.2\%) & 218 (27.1\%)\\
\hspace{1em}Other & 36 (9.3\%) & 30 (7.2\%) & 66 (8.2\%)\\
\addlinespace[0.3em]
\textbf{Accelerated} & 95 (24.5\%) & 94 (22.6\%) & 189 (23.5\%)\\
\addlinespace[0.3em]
\textbf{EIP}
 & 28 (7.2\%) & 29 (7.0\%) & 57 (7.1\%)\\
\addlinespace[0.3em]
\textbf{Gifted}
 & 75 (19.3\%) & 75 (18.0\%) & 150 (18.7\%)\\
\addlinespace[0.3em]
\textbf{IEP}
 & 37 (9.5\%) & 41 (9.9\%) & 78 (9.7\%)\\
\addlinespace[0.3em]
\textbf{ESOL}
 & 38 (9.8\%) & 34 (8.2\%) & 72 (9.0\%)\\
\addlinespace[0.3em]
\multicolumn{4}{l}{\textbf{Days Absent (6th Grade)}}\\
\hspace{1em}Median [IQR] & 2.46 [3.87] & 2.58 [3.74] & 2.52 [3.83]\\
\addlinespace[0.3em]
\multicolumn{4}{l}{\textbf{Average Time on Task during Pretest
 (Minutes)}}\\
\hspace{1em}Median [IQR] & 1.04 [0.895] & 1.06 [0.880] & 1.05 [0.903]\\
\addlinespace[0.3em]
\multicolumn{4}{l}{\textbf{Math Anxiety}}\\
\hspace{1em}Mean (SD) & 13.5 (5.64) & 13.5 (5.74) & 13.5 (5.69)\\
\addlinespace[0.3em]
\multicolumn{4}{l}{\textbf{Math Self-Efficacy}}\\
\hspace{1em}Mean (SD) & 17.5 (4.81) & 17.7 (4.72) & 17.6 (4.76)\\
\addlinespace[0.3em]
\multicolumn{4}{l}{\textbf{Perceptual Sensitivity}}\\
\hspace{1em}Mean (SD) & 7.67 (2.91) & 7.66 (2.92) & 7.66 (2.91)\\
\addlinespace[0.3em]
\multicolumn{4}{l}{\textbf{\% Correct w/o Feedback}}\\
\hspace{1em}Mean (SD) & NA & 65.9 (16.1) & 65.9 (16.1)\\
\addlinespace[0.3em]
\multicolumn{4}{l}{\textbf{\% Partial Feedback}}\\
\hspace{1em}Mean (SD) & NA & 23.2 (7.93) & 23.2 (7.93)\\
\addlinespace[0.3em]
\multicolumn{4}{l}{\textbf{\% Bottom Out}}\\
\hspace{1em}Mean (SD) & NA & 10.9 (12.2) & 10.9 (12.2)\\
\addlinespace[0.3em]
\multicolumn{4}{l}{\textbf{Posttest (i.e. State Standardized Test Score)}}\\
\hspace{1em}Mean (SD) & 9.04 (1.01) & 9.05 (0.994) & 9.05 (0.999)\\
\bottomrule
\end{tabular}
}
\end{table}

\begin{table}[pt!]
\centering
\caption{Estimated coefficients and standard errors from models predicting $\etat$ as a function of covariates. 
}
\vspace{-0.5em}
\label{tab:usageReg}
\begin{threeparttable}
{\small
\begin{tabular}{lrrrrrr}
  \toprule
  & Rasch &  & 2PL &  & GRM &  \\
\midrule
Male & --0.07 * & (0.03) & --0.09 & (0.05) & --0.06 & (0.05) \\
  Hispanic/Latino & --0.01 & (0.03) & --0.01 & (0.06) & 0.01 & (0.07) \\
  Asian & 0.09 * & (0.04) & 0.1 & (0.07) & 0.09 & (0.07) \\
  Other & --0.02 & (0.03) & --0.04 & (0.06) & --0.02 & (0.06) \\
  Accelerated & 0.11 * & (0.05) & 0.1 & (0.09) & 0.05 & (0.1) \\
  5th Grd State Test & 0.37 * & (0.05) & 0.5 * & (0.09) & 0.53 * & (0.09) \\
  Pretest & 0.15 * & (0.05) & 0.2 * & (0.09) & 0.18 & (0.09) \\
  EIP & --0.06 * & (0.03) & --0.08 & (0.06) & --0.12 * & (0.06) \\
  ESOL & 0.01 & (0.04) & 0.02 & (0.06) & 0.04 & (0.06) \\
  Gifted & 0.05 & (0.03) & 0.07 & (0.06) & 0.07 & (0.06) \\
  IEP & 0.02 & (0.03) & -0 & (0.05) & 0.01 & (0.05) \\
  $\sqrt{6^{th}\mbox{ Grd Days Absent}}$ & --0.02 & (0.03) & --0.03 & (0.06) & --0.04 & (0.06) \\
  Log(Pretest Avg. Time) & 0.07 * & (0.03) & 0.08 & (0.06) & 0.1 & (0.06) \\
  Math Anxiety & --0.06 & (0.04) & --0.08 & (0.07) & --0.06 & (0.07) \\
  Math Self-Efficacy & 0.04 & (0.03) & 0.04 & (0.07) & 0.07 & (0.07) \\
  Perceptual Sensitivity & 0.14 * & (0.04) & 0.12 & (0.08) & 0.12 & (0.08) \\
   \hline
$\sigma_\eta$ & 0.49 &  & 1 &  & 1 &  \\
  $R^2$ & 0.8 &  & 0.6 &  & 0.61 &  \\
\bottomrule
\end{tabular}
}

\vspace{0.5em}
\begin{tablenotes}
    \linespread{1} \footnotesize
    \item \textit{Note}. * Central 95\% credible interval excludes 0. Teacher effects are omitted.
\end{tablenotes}
\end{threeparttable}
\vspace{-1.5em}
\end{table}

\begin{table}
\centering
\caption{Structural Parameters from outcome submodels. 
}
\label{tab:outcomeReg}
\vspace{-0.5em}
\begin{threeparttable}
{\small
\begin{tabular}{lrrrrrr}
  \hline
  & Rasch &  & 2PL &  & GRM &  \\
   \hline
(Intercept) & 9.05 * & (0.03) & 9.04 * & (0.03) & 9.03 * & (0.03) \\
  $\omega$ & 0.46 * & (0.06) & 0.35 * & (0.04) & 0.32 * & (0.05) \\
  $\tau_0$ & --0.01 & (0.04) & 0.02 & (0.04) & 0.03 & (0.04) \\
  $\tau_1$ & 0.03 & (0.04) & 0.06 & (0.04) & 0.04 & (0.04) \\
  Male & 0.04 * & (0.02) & 0.03 & (0.02) & 0.03 & (0.02) \\
  Hispanic/Latino & -0 & (0.02) & --0.01 & (0.02) & --0.01 & (0.02) \\
  Asian & --0.04 & (0.03) & --0.02 & (0.03) & --0.01 & (0.03) \\
  Other & --0.02 & (0.02) & --0.02 & (0.02) & --0.03 & (0.02) \\
  Accelerated & 0.09 * & (0.04) & 0.11 * & (0.03) & 0.13 * & (0.03) \\
  5th Grd State Test & 0.24 * & (0.04) & 0.27 * & (0.04) & 0.28 * & (0.04) \\
  Pretest & 0.05 & (0.04) & 0.06 & (0.03) & 0.07 * & (0.03) \\
  EIP & --0.01 & (0.02) & --0.01 & (0.02) & --0.01 & (0.02) \\
  ESOL & --0.04 & (0.02) & --0.04 & (0.02) & --0.04 & (0.02) \\
  Gifted & 0.05 * & (0.02) & 0.05 * & (0.02) & 0.05 * & (0.02) \\
  IEP & --0.05 * & (0.02) & --0.04 * & (0.02) & --0.05 * & (0.02) \\
  $\sqrt{6^{th}\mbox{ Grd Days Absent}}$ & --0.02 & (0.02) & --0.02 & (0.02) & --0.02 & (0.02) \\
  Log(Pretest Avg. Time) & --0.05 * & (0.02) & --0.04 * & (0.02) & --0.04 * & (0.02) \\
  Math Anxiety & -0 & (0.02) & --0.01 & (0.02) & --0.02 & (0.02) \\
  Math Self-Efficacy & 0.03 & (0.02) & 0.04 & (0.02) & 0.03 & (0.02) \\
  Perceptual Sensitivity & --0.01 & (0.03) & 0.02 & (0.03) & 0.03 & (0.03) \\
   \hline
$\sigma_Y$ & 0.45 &  & 0.44 &  & 0.45 &  \\
  $R^2$ & 0.53 &  & 0.62 &  & 0.63 &  \\
   \hline
  \end{tabular}
}

\vspace{0.5em}
\begin{tablenotes}
    \linespread{1} \footnotesize
    \item \textit{Note}. * Central 95\% credible interval excludes 0. Covariate coefficients and teacher fixed-effects omitted.
\end{tablenotes}
\end{threeparttable}
\vspace{-1.5em}
\end{table}

\clearpage

\subsection{Ability and Difficulty within the Dataset}

Using data from the ``Instant'' condition, we fit the following model using the \texttt{lme4} package in \texttt{R} \cite{lme4,rcite}:

\begin{equation}\label{eq:raschApp}
Pr(Y_{ij}=1|\bm{\eta_T},\bm{d})=logit^{-1}\left(\alpha_0+\eta_{Ti}-d_j\right)
\end{equation}
where $Y_{ij}=1$ if student $i$ answered problem part $j$ without making any errors or requesting a hint. Student ability parameters $\eta_T\sim\mathcal{N}(0,\sigma_\eta)$ and problem difficulty parameters $d\sim\mathcal{N}(0,\sigma_d)$.
\pagebreak

\subsubsection{Problem Difficulty Throughout the Curriculum}
There were nine sections, each containing a different number of problems.
Students worked on all of the problems in the same order.

The following is a plot of estimated $d_j$ from model \eqref{eq:raschApp} against problem order within each of the nine problem sets:
\begin{center}
\includegraphics[width=0.95\textwidth]{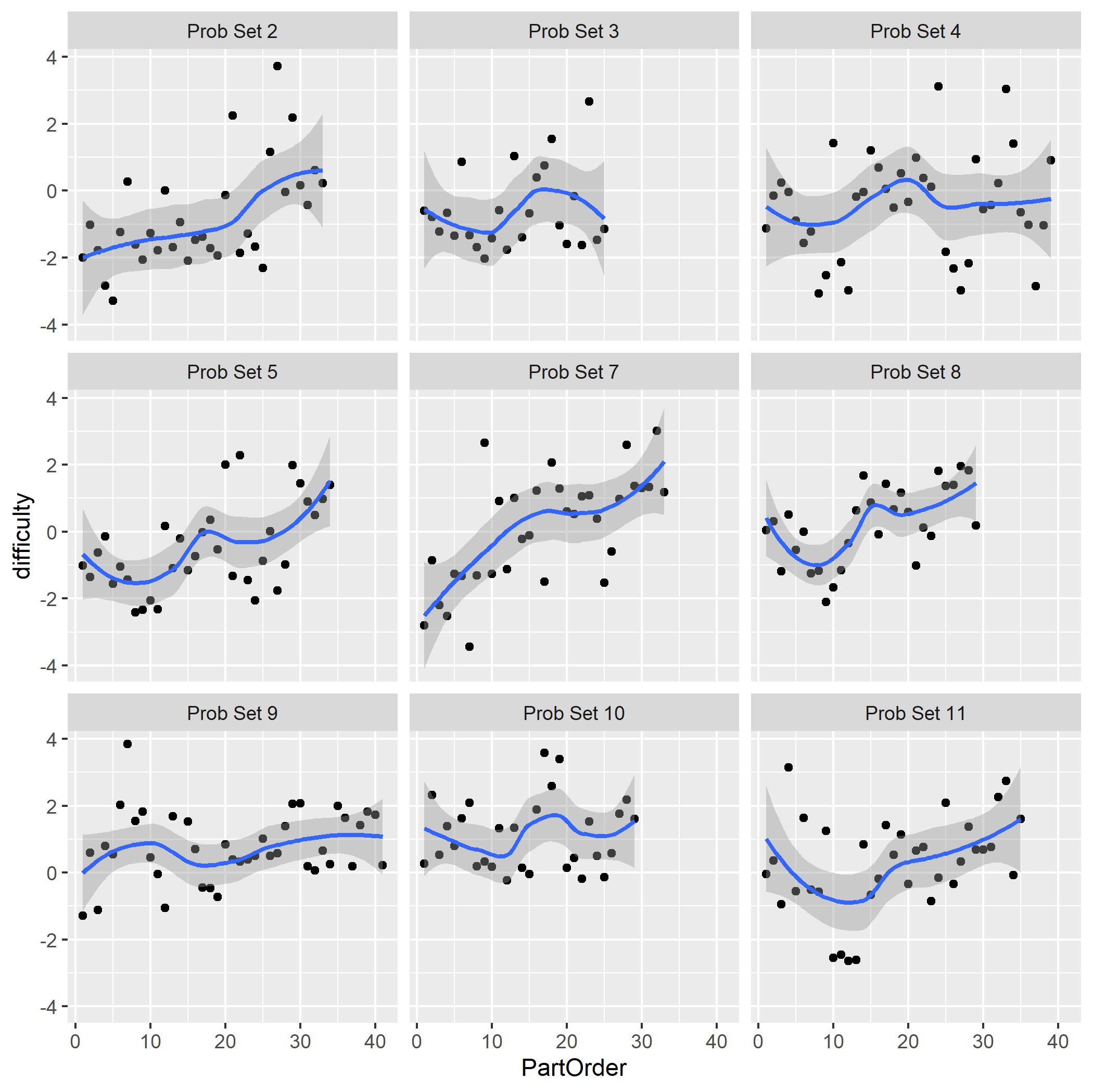}
\end{center}

\pagebreak

The following plots the average estimated $d_j$ within each of the nine problem sets:
\begin{center}
\includegraphics[width=0.8\textwidth]{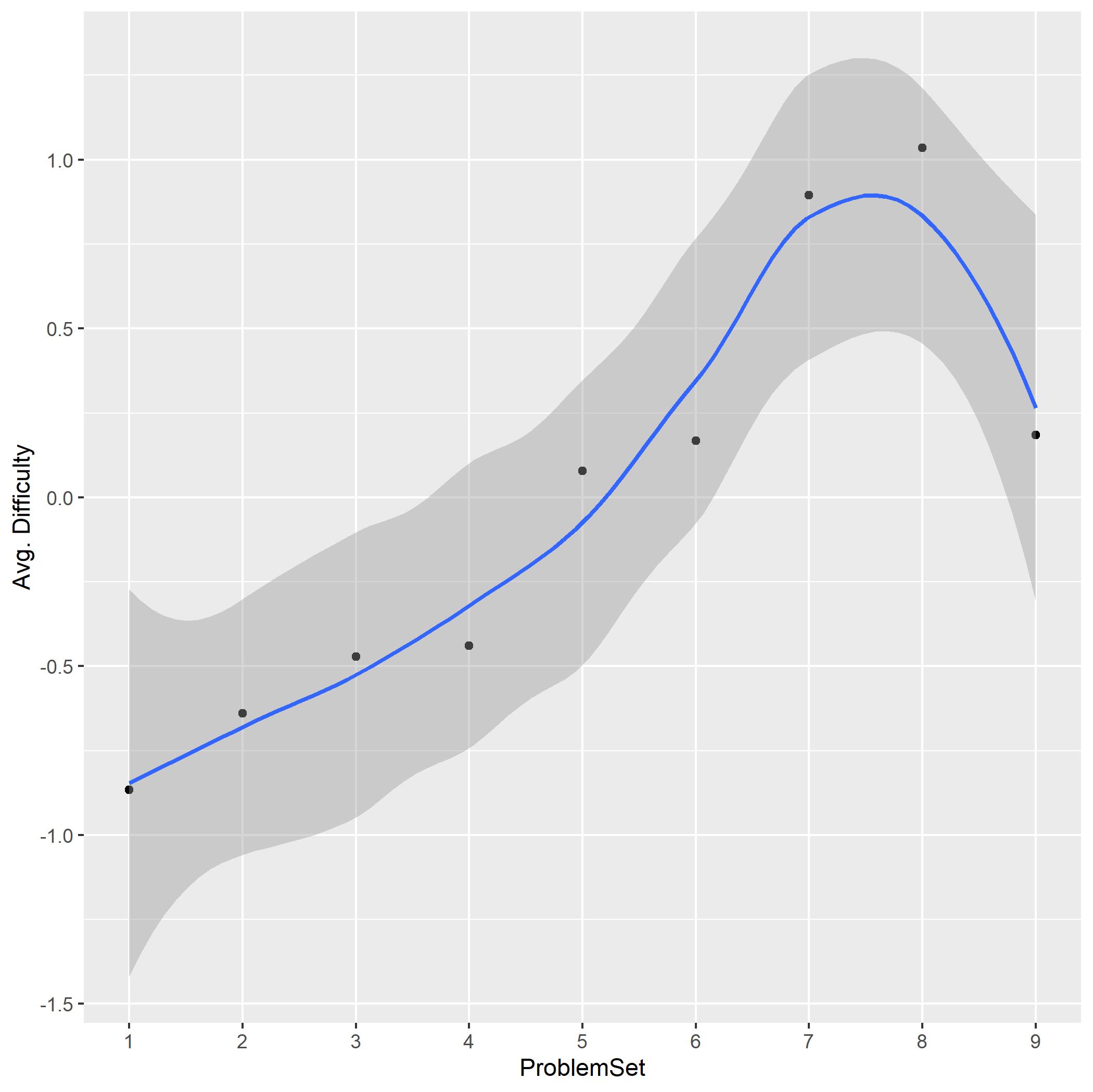}
\end{center}
\pagebreak

\subsubsection{Student Ability by Problem}
For each problem in the curriculum, we estimated the average ability for students attempting that problem as $\bar{\eta_T}_j=\sum_{i=1}^n W_{ij}\hat{\eta}_{Ti}/\sum_{i=1}^n W_{ij}$, where $n$ is the total number of students in the Instant condition and $W_{ij}=1$ if student $i$ worked on problem $j$ and 0 otherwise.

The following plots $\bar{\eta_T}$ against problem order within each of the nine problem sets:
\begin{center}
\includegraphics[width=0.95\textwidth]{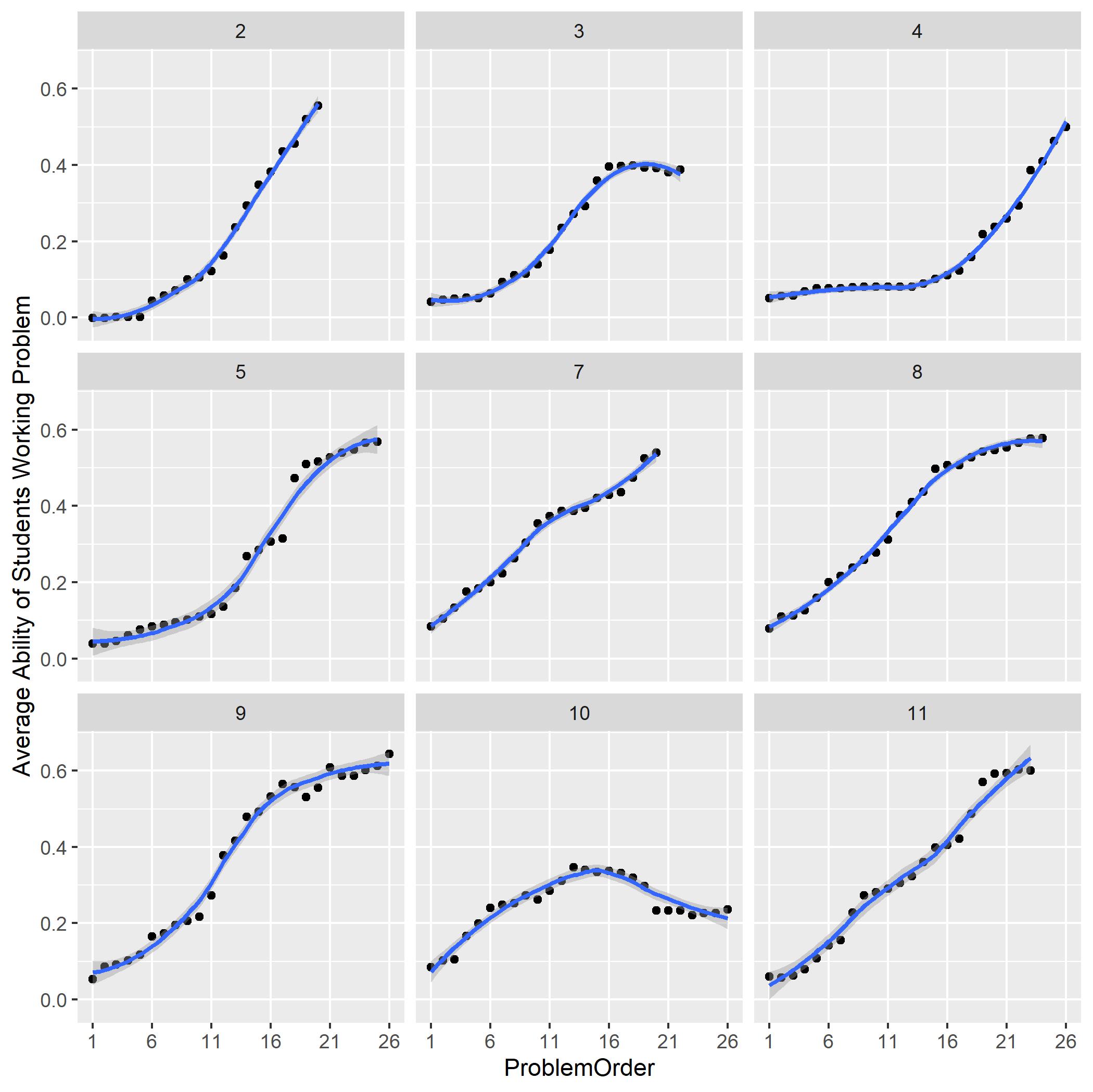}
\end{center}
\pagebreak
The following plots the average of estimated $\eta_T$ for all students who attempted any problem in each of the nine problem sets:

\begin{center}
\includegraphics[width=0.8\textwidth]{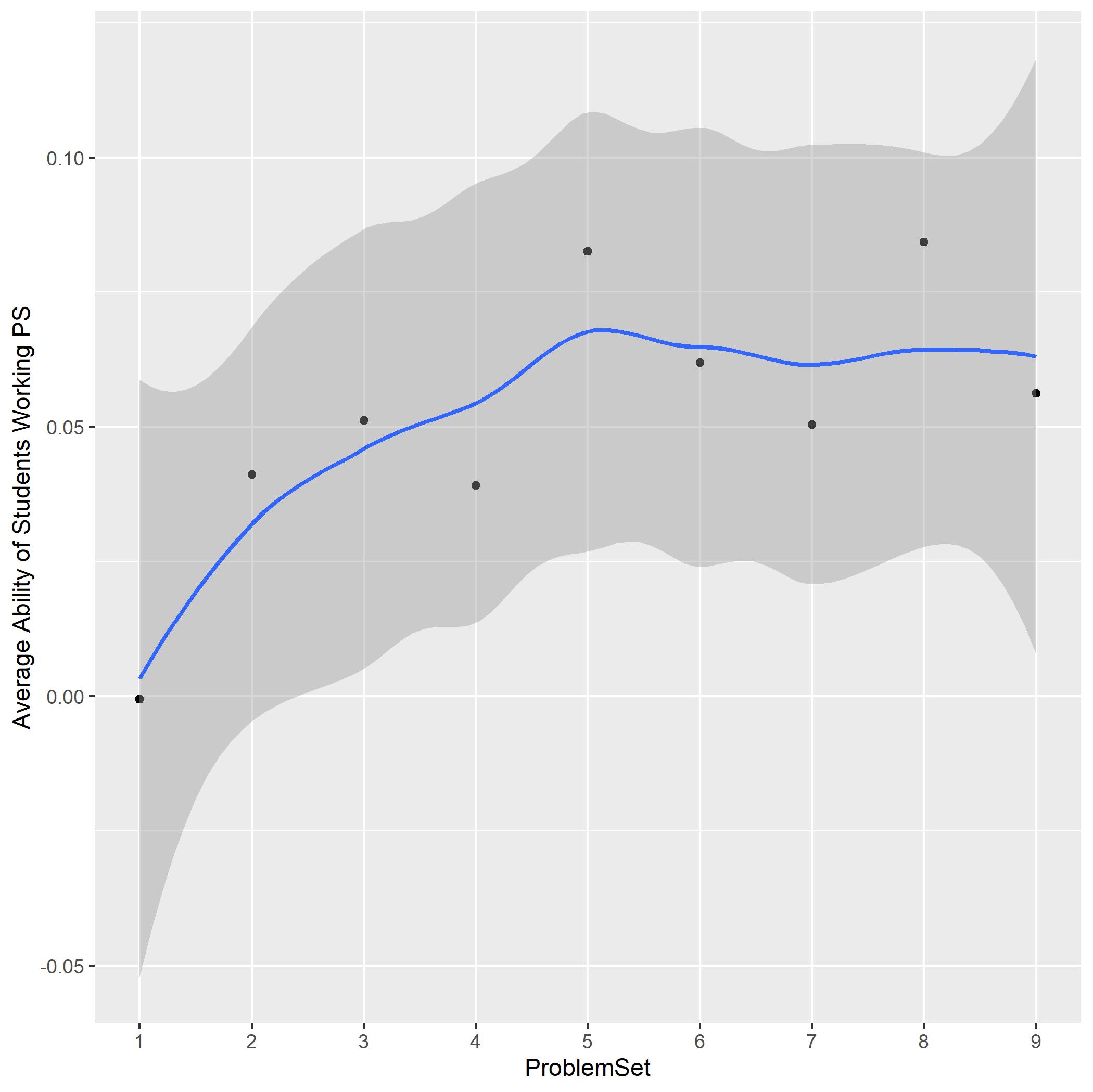}
\end{center}

\end{document}